\documentclass[11pt]{article} 
\makeatletter\renewcommand{\section}{\@startsection
	{section}{1}{\z@}{-3.5ex plus -1ex minus
		-.2ex}{2.3ex plus .2ex}{\bf }}
\makeatletter\renewcommand{\subsection}{\@startsection{subsection}{2}{\z@}{-3.25ex
		plus -1ex minus
		-.2ex}{1.5ex plus .2ex}{\it }}
\makeatletter\renewcommand{\subsubsection}{\@startsection{subsubsection}{3}{-2.45ex}{-3.25ex
		plus -1ex minus -.2ex}{1.5ex plus .2ex}{\it }}

\renewcommand{\theequation}{\thesection.\arabic{equation}}
\makeatletter \@addtoreset{equation}{section}
\usepackage[table,xcdraw]{xcolor}
\renewenvironment{thebibliography}[1]
{\baselineskip=16pt plus 2pt minus 1pt
	\section*{\large\refname
		\@mkboth{\MakeUppercase\refname}{\MakeUppercase\refname}}%
	\list{\@biblabel{\@arabic\c@enumiv}}%
	{\settowidth\labelwidth{\@biblabel{#1}}%
		\leftmargin\labelwidth
		\advance\leftmargin\labelsep
		\@openbib@code
		\usecounter{enumiv}%
		\let\p@enumiv\@empty
		\renewcommand\theenumiv{\@arabic\c@enumiv}}%
	\sloppy
	\clubpenalty4000
	\@clubpenalty \clubpenalty
	\widowpenalty4000%
	\sfcode`\.\@m}

\let\fn\footnote
\renewcommand{\footnote}[1]{\linespread{1.1}\fn{#1}\linespread{1.29}}

\usepackage{multirow}
\usepackage{epsfig} 
\usepackage{verbatim}
\usepackage{graphicx}
\usepackage{tikz}
\usetikzlibrary{trees,er,snakes,shapes,mindmap}
\usepackage{hyperref}
\usepackage{easyfig}
\usepackage{bbm}
\usepackage{cite}
\usepackage{slashed}
\usepackage{physics}
\usepackage{amsmath}
\usepackage{amsfonts}
\usepackage{amssymb}
\usepackage{braket}
\usepackage{mathrsfs}
\usepackage{bm}
\usepackage[section]{placeins}
\usepackage{placeins}
\usepackage{mathtools}
\usepackage[utf8x]{inputenc}
\usepackage{booktabs}
\usepackage{caption}

\def\tyng(#1){\hbox{\tiny$\yng(#1)$}}

\newcommand{\del}{\partial}

\let\Oldsection\section
\renewcommand{\section}{\FloatBarrier\Oldsection}

\let\Oldsubsection\subsection
\renewcommand{\subsection}{\FloatBarrier\Oldsubsection}

\let\Oldsubsubsection\subsubsection
\renewcommand{\subsubsection}{\FloatBarrier\Oldsubsubsection}

\allowdisplaybreaks
\newcommand{\be}{\begin{equation}}
\newcommand{\ee}{\end{equation}}
\newcommand{\beq}{\begin{equation}}
\newcommand{\eeq}{\end{equation}}
\newcommand{\bea}{\begin{array}}
	\newcommand{\ea}{\end{array}}
\newcommand{\beqa}{\begin{eqnarray}}
\newcommand{\eeqa}{\end{eqnarray}}
\newcommand{\beqar}{\begin{eqnarray}}
\newcommand{\eeqar}{\end{eqnarray}}
\newcommand{\nn}{\nonumber}
\def\la{\langle}
\def\ra{\rangle}
\def\half{\textstyle{1\over 2}}
\def\bz{{\bar z}}
\DeclareMathOperator{\Ree}{\mathfrak{Re}}

\begin{document}
\fontfamily{bch}\fontsize{11pt}{15pt}\selectfont
	\begin{titlepage}
		\begin{flushright}
		\end{flushright}
		
		\vskip 2 em
		
		\begin{center}
{\Large \bf Magnetic Field and Curvature Effects on Pair Production}\\
~\\
{\Large \bf I: Scalars and Spinors}
			
			
			\vskip 1.5cm
			
			\centerline{$ \text{\large{\bf{D. Karabali}}}^{a} $, $ \text{\large{\bf{S. K\"{u}rk\c{c}\"{u}o\v{g}lu}}}^{b,c} $, $ \text{\large{\bf{V.P.Nair}}}^{c} $}
			
			\vskip 0.5cm
			\centerline{\sl $^a$ Department of Physics and Astronomy}
			\centerline{\sl Lehman College of the CUNY, Bronx, NY 10468, USA}
			\vskip 1em
			\centerline{\sl $^b$ Middle East Technical University, Department of Physics,}
			\centerline{\sl Dumlupinar Boulevard, 06800, Ankara, Turkey}
			\vskip 1em
			\centerline{\sl $^c$ Physics Department, City College of the CUNY   }
			\centerline{\sl New York, NY 10031, USA}
			\vskip 1em
 \vskip .26cm
\begin{tabular}{r l}
E-mail:&\!\!\!{\fontfamily{cmtt}\fontsize{11pt}{15pt}\selectfont dimitra.karabali@lehman.cuny.edu}\\
&\!\!\!{\fontfamily{cmtt}\fontsize{11pt}{15pt}\selectfont kseckin@metu.edu.tr}\\
&\!\!\!{\fontfamily{cmtt}\fontsize{11pt}{15pt}\selectfont vpnair@ccny.cuny.edu}
\end{tabular}
			
		\end{center}
		
		\vskip 4 em
		
		\begin{quote}
			\begin{center}
				{\bf Abstract}
\end{center}
			
\vskip 1em

The pair production rates for spin-zero and spin-$\half$ particles are calculated 
on spaces of the form $M \times {\mathbb R}^{1,1}$
with $M$ corresponding to ${\mathbb R}^2$ (flat), $T^2$ (flat, compactified), $S^2$
(positive curvature) and $H^2$ (negative curvature),
with and without a background magnetic field on $M$.
The motivation is to elucidate the effects of curvature and background
magnetic field.
Contrasting effects for positive and negative curvature on the two cases of spin 
are obtained. For positive curvature, we find enhancement
for spin-zero  and suppression for spin-$\half$, with the opposite effect for
negative curvature.

\vskip 1em
			
\vskip 5pt

\end{quote}
		
\end{titlepage}
\setcounter{footnote}{0}
\pagestyle{plain} \setcounter{page}{2}
	
\newpage

\section{Introduction}

The Schwinger process of pair production of charged particles in a uniform electric field
has long remained a topic of research interest since the original calculation
\cite{Schwinger:1951nm}.
It has been analyzed by different techniques, extended, at least to some extent, to the case of 
spacetime-dependent fields \cite{dunne}, to the dual situation of the production of magnetic monopole pairs in a magnetic
field \cite{Affleck:1981ag},
 etc. Mathematically similar calculations also apply to the case
of spacetime curvature as the agent of particle production \cite{qft-curved}.
A key feature of the Schwinger process is that, while there is a nonzero probability
for pair production for all values of the electric field, there is a suppression effect
due to the mass $m$ of the charged particles, of the form $e^{-m^2 \pi /E}$ where 
$E$ denotes the electric field.
This is essentially due to the fact that a pair needs a minimal energy of $2 m$ to be
liberated as free particles from the vacuum.
This, of course, immediately brings up the question of whether or how the process 
can be enhanced
in a situation where the liberated pair are not free particles, but in a bound state, so that
the binding energy effectively reduces the suppression due to the mass.
Pair production where the created pair ends up in bound states has not 
yet been investigated in detail and this provides the general setting and motivation
 for the present work.

Relevant to this issue is the fact that
the electrostatic potential energy will have opposite signs for the two members
of a charged pair and hence
additional electric fields do not necessarily provide a good example.
One needs a situation for which both members of the pair can be in bound states.
For example, a whole background of protons and antiprotons, in addition to
the constant background electric field, can provide binding centers
for the created electrons and positrons and could lead to a situation of enhanced
pair production. However, this problem is practically intractable.

Gravitational fields can provide charge-symmetric binding for the created pair, so 
the analysis of the Schwinger process in a background of nonzero curvature
is one case which should be interesting\cite{Mottola:1984ar}. 
In a background magnetic field, we get Landau
levels for both types of charges, so this is another case worth exploring
in more detail \cite{Kim:2003qp}. The spin of the particle also affects pair production in a nontrivial way.
With an additional magnetic field, the Zeeman coupling of the spin becomes important
in determining the energy levels. For gravitational backgrounds as well, there is
a Zeeman-like coupling of spin to curvature which can affect the process.
Recall that Schwinger's original calculation involved the use of 
the spacetime trajectories of
particles. While spinless particles follow geodesics in a gravitational background, spinning
particles follow the Mathisson-Papapetrou trajectories
due to the curvature-spin coupling. Thus even within Schwinger's 
calculational framework, we can see that there will be interesting spin
effects when there is an additional
magnetic field or spacetime curvature present.

With this motivation and background, in this paper and in an accompanying paper,
we consider the pair production of particles of spins zero, $\half$ and $1$
in a background with both electric and magnetic fields.
For the geometric background, we will consider manifolds of the form
$M \times {\mathbb R}^{1,1}$, where
$M$ will be taken to be
${\mathbb R}^2$ (so that the total space is flat Minkowski spacetime),
and $T^2$, $S^2$ and $H^2$ in turn.
The uniform electric field is taken to be in the $ {\mathbb R^{1,1}}$ components
(which have metrical signature $+ ,-$) ,
with a magnetic field on ${\mathbb R}^2$, $T^2$, $S^2$ and $H^2$.
Here $T^2$ is the two-dimensional torus,
$S^2$ is the two-sphere and $H^2$ is the two-dimensional hyperboloid.
These can be viewed as ${\mathbb R}^2$ modulo a square lattice,
$SU(2)/U(1)$ and $SL(2, {\mathbb R})/U(1)$ respectively,
so that a group-theoretic analysis of the effective action is possible.
(There are also some features which make this analysis
interesting in its own right.)
We have chosen $T^2$, $S^2$ and $H^2$ to exemplify
the cases of flat but compact, positively curved,
and negatively curved spaces, respectively.

In this paper, we will consider spins zero and $\half$. The case of 
spin $1$ needs a more elaborate discussion. The only consistent
approach to spin-$1$ particles is to treat them as part of
a nonabelian gauge field. The Yang-Mills action will then determine the
correct Zeeman coupling as well as the spin-curvature coupling.
For these fields, the correct counting of the physical
degrees of freedom is also nontrivial and will require a BRST analysis.
Finally, it is known that, in the nonabelian case,
 there is an instability even for background fields of the purely magnetic type,
even without an electric-type field. This has to be accounted for
in the formalism. An interesting postscript to some of the older
attempts to understand confinement will also result from our analysis.
For all these reasons, we will discuss the spin-$1$ case
in a separate paper.

This paper is organized as follows. In section 2 we discuss flat space,
pointing out additional enhancement and suppression effects with a 
background magnetic field. This is followed by sections
discussing the torus, two-sphere and hyperboloid in sequence.
The corresponding results are compared to the case of flat Minkowski
geometry.
The paper concludes with a discussion and an appendix 
which is a short resum\'e of results on $SL(2, {\mathbb R})$
relevant to our analysis.

\section{Pair production with electric and magnetic fields: Flat space and spins zero and $\half$}

\subsection{Scalar Field, spin zero}

As mentioned in the introduction, our aim is to consider the effect of 
Landau levels as well as curvature on pair 
production for particles of different spins in a uniform electric field. 
The simplest approach to calculating this is to consider two orthogonal
magnetic fields in Euclidean space, for which we have an explicit solution 
in terms of the Landau levels.
It is then straightforward to calculate the effective action and then continue
to Minkowski space, with one of the magnetic fields continuing to the electric field.

This strategy is easily illustrated  for the case of a scalar charged field $\phi$. The Euclidean
action can be taken as
\be
S = \int d^4x \left[ (D_\mu \phi )^*  (D^\mu\phi) + m^2 \, \phi^* \phi \right] 
\label{rev1}
\ee
where $D_\mu$ denotes the covariant derivative as usual.
The corresponding effective action, upon integrating out the field $\phi$, is
\beq
\Gamma = \mbox{Tr} \, \log \, (- D^2 + m^2) \label{rev2}
\eeq
We must evaluate this determinant for constant magnetic fields, which will be taken
to be $F_{12} = B_1$ in the $(1,2)$-plane and $F_{34} = B_2$ in the $(3,4)$-plane.
Using $\Pi_\mu = -i D_\mu= -i \del_\mu - A_\mu$, we have
$[ \Pi_1 , \Pi_2 ] = i (\del_1 A_2 - \del_2 A_1) = i \, B_1$, 
$[ \Pi_3 , \Pi_4 ] = i (\del_3 A_4 - \del_4 A_3) = i \, B_2$, 
We thus have two sets of canonically conjugate operators
$(\Pi_1 / \sqrt{B_1},  \Pi_2 /\sqrt{B_1} )$, $(\Pi_3 / \sqrt{B_2},  \Pi_4 /\sqrt{B_2} )$.
Writing $\Pi_\mu \Pi^\mu$ in terms of these, we find  the eigenstates and eigenvalues,
\beq
\left[ (\Pi_1^2 + \Pi_2^2 ) + (\Pi_3^2 + \Pi_4^2 ) \right]\, \psi_{n_1, n_2, \alpha}
=  \left[ (2 n_1 +1) B_1 + (2 n_2 +1) B_2 \right]\, \psi_{n_1, n_2, \alpha}
\label{rev3}
\eeq
where $n_1$, $n_2$ are either zero or positive integers. The subscript $\alpha$ 
on the wave functions denotes degeneracy.
Using this result we find the effective action
\begin{align}
\Gamma &= - \mbox{Tr} \int_\epsilon^\infty  {ds \over s} \, e^{-s \, (-D^2 + m^2)} = - \int d^4x\, {ds \over s}~ \la x \vert\, e^{-s \, (-D^2 + m^2) } \,\vert x\ra\nonumber\\
&= - \int_\epsilon^\infty {ds \over s} ~\int d^4x \sum_{n_1, n_2, \alpha}
\psi^*_{n_1, n_2,\alpha}(x)\, \psi_{n_1, n_2, \alpha}(x) \, \,
e^{-s (2n_1 +1) B_1 - s (2 n_2 +1 ) B_2 - s m^2 }\nonumber\\
&= - \int_\epsilon^\infty  {ds \over s}\, d^4x\, {B_1 B_2 \over (2\pi)^2} \sum_{n_1, n_2}
e^{-s (2n_1 +1) B_1 - s (2 n_2 +1 ) B_2 - s m^2}
\label{rev4}
\end{align}
Here $\epsilon $ is a small positive real number which 
can be taken to be zero after renormalization.
We have also used the normalization condition
$\int d^4x \,\psi^*_{n_1, n_2, \alpha}(x) \psi_{n_1, n_2, \alpha}(x) = 1$ and the fact
that the degeneracy of the Landau levels is
given by the factor
$(B_1 /2 \pi) dx_1 dx_2 \times (B_2 /2 \pi) dx_3 dx_4$.
In the present case, we can actually carry out the summations
in (\ref{rev4})
to obtain a closed form formula,
\beq
\Gamma = - {1\over 16\pi^2} \int_\epsilon^\infty {ds \over s^3} \int d^4x\,e^{- s m^2} \, \left( {s B_1 \over \sinh sB_1}\right)\,
\left( {s B_2 \over \sinh sB_2}\right)
\label{rev5}
\eeq
There are divergences at $s = \epsilon$ as $\epsilon \rightarrow 0$; these correspond to ultraviolet divergences and have to be subtracted out, for which one can use the expansion
\beq
{ s B \over \sinh sB} \simeq 1 - {(s B )^2 \over 6} ~+~ {\cal O}(s^4)
\label{rev6}
\eeq
This subtraction procedure has to be carried out to obtain the real part of the effective
action. However, our focus here is on the decay rate of the vacuum state
due to particle production. There are no ultraviolet divergences for this.
Also, for the case of the sphere and the hyperboloid, explicit summation will not be possible.
So, to calculate the decay rate along the lines which generalize to the curved manifold cases,
we go back to (\ref{rev4}) and carry out the summation
over $n_2$ to write
\beq
\Gamma = - {1\over 8 \pi^2} \int_\epsilon^\infty {d s\over s^2} d^4x\, \left( {s B_2 \over \sinh sB_2}\right)\,
 \left[ {B_1}
\sum_{n_1} e^{- s ( m^2 + (2 n_1 + 1)B_1 )}\right]
\label{rev7}
\eeq
We consider the continuation of this result to Minkowski space by using
$x_4 \rightarrow i x_0$, $B_2 \rightarrow -i E $. Further,
the continuation of $-\Gamma$ is to be identified as $i S_{\rm eff}$,
with $\braket{0|0} = e^{i S_{\rm eff}}$.
We are thus interested in the real part of $i S_{\rm eff}$.
From (\ref{rev7}),
\beq
i S_{\rm eff} = {1 \over 8 \pi^2} \int i \,d^4x~ \int_\epsilon^\infty {ds \over s^2} 
{s E \over \sin s E} \,  \left[ {B_1}
\sum_{n_1} e^{- s ( m^2 + (2 n_1 + 1)B_1 )}\right]
\label{rev8}
\eeq
This expression has singularities at $s E = n \, \pi$, $n = 1, 2, \cdots$.
(There is no singularity at $n =0$,  or $s =0$ since the
integration starts at $s = \epsilon$. To put another way,
the $s =0$ singularity is subtracted out
via renormalization.)
The imaginary part of $S_{\rm eff}$ (i.e. the real part of $i S_{\rm eff}$)
arises from going around the singularities in doing the $s$-integration.
Near $s E = n \pi$, we write
$s = (n \pi /E) + z$, $\sin s E = \sin (n \pi + E z) \simeq (-1)^n E z$.
We then get, for the contribution from a small semicircle around these points,
\begin{align}
\Ree (i S_{\rm eff} )& = {i \over 8 \pi^2} \int d^4x\,  \sum_{n=1}^\infty  \, (-1)^n (E/n \pi ) 
\int_\pi^0 {dz \over z} \,  \left[ {B_1}
\sum_{n_1} e^{- (n \pi/ E) ( m^2 + (2 n_1 + 1)B_1 )}\right]\nonumber\\
&= \int d^4x \, {E \over 8 \pi^2} \sum_{n=1}^\infty  {(-1)^n \over n} 
 \left[ {B_1}
\sum_{n_1} e^{- (n \pi/ E) ( m^2 + (2 n_1 + 1)B_1 )}\right]
\label{rev9}
\end{align}
If we carry out the summation over $n_1$ and then take the limit $B_1 \rightarrow 0$,
this becomes
\beqar
\Ree (i S_{\rm eff}) &=& - \int d^4x\, {E^2 \over 16 \pi^3}  \sum_{n=1}^\infty {(-1)^{n+1} \over n^2} e^{- m^2 (n \pi/E)}
\label{rev10}\\
&=& - \int d^4x\, {E^2 \over 192 \pi}, \hskip .2in
{\rm as}~ m \rightarrow 0\nonumber
\eeqar
In the last line we have used the fact that $\sum_{n=1}^\infty (-1)^{n+1} /n^2 = \pi^2 /12$.
This result (\ref{rev10}) in agreement with the standard Schwinger calculation with just the electric field
in flat space.

\subsection{Dirac Field, spin $\half$}

In the case of the Dirac field, the effective action is given by
\begin{align}
\Gamma &= - \mbox{Tr} \log \left( i \gamma\cdot \Pi + m\right)\nonumber\\
&= - {1\over 2} \mbox{Tr} \log  \left( i \gamma\cdot \Pi + m\right)  \left( -i \gamma\cdot \Pi + m\right)\nonumber\\
&= -{1\over 2} \mbox{Tr} \log \left[ \Pi^2 + m^2 + \left( {1\over 2}[\gamma_\mu, \gamma_\nu]\right)
\left({1\over 2} [ \Pi_\mu, \Pi_\nu]\right) \right]
\label{rev11}
\end{align}
Our choice of Euclidean $\gamma$-matrices in terms of the Pauli matrices $\sigma_i$ 
is
\beq
\gamma_i = \left[ \begin{matrix} 0& i \sigma_i\\
-i \sigma_i &0\\ \end{matrix} \right], \quad \quad
\gamma_4 = \left[ \begin{matrix} 0& 1\\
1&0\\ \end{matrix} \right]
\label{rev11b}
\eeq
This explicit representation of $\gamma_\mu$  is not needed for most calculations.
The operator in (\ref{rev11}) is the same as before, except for the Zeeman term. 
With the background fields as before, there are four distinct sets of eigenvalues
(corresponding to the sign combinations
$++$, $+-$, $-+$ and $--$)
given by
\beq
\Pi^2 + \left( {1\over 2}[\gamma_\mu, \gamma_\nu]\right)
\left({1\over 2} [ \Pi_\mu, \Pi_\nu]\right) 
= (2 n_1 +1) B_1 + (2n_2 + 1) B_2 \pm B_1 \pm B_2
\label{rev11a}
\eeq
Unlike the spin zero case,
there is a mode with eigenvalue equal to zero, where the Zeeman term cancels the
zero point contribution from $\Pi^2$.
The degeneracy, as before,
is given by $d^4x (B_1 B_2 /4\pi^2)$. Thus
carrying out the summation over $n_2$ as before, we find
\beq
\Gamma = {1\over 2} \int {d s \over s} d^4x\, {B_1 B_2 \over 4 \pi^2} \,\coth s B_2\,
\sum_{n_1} \left( e^{- s (m^2 + 2 n_1 B_1 ) } + e^{ - s (m^2 + (2 n_1 + 2)B_1 )}
\right) 
\label{rev12}
\eeq
Continuing this expression to Minkowski space,
we find the real part of $i S_{\rm eff}$ as
\beq
\Ree (i S_{\rm eff}) = - \int d^4x\,{ E \over 8 \pi^2} 
\sum_{n=1}^\infty {1\over n} \left[ B_1 e^{- s m^2} \sum_{n_1} \left( 1 + e^{- 2 s B_1}\right)
e^{- 2 s n_1 B_1} \right]_{s= (n \pi /E)}
\label{rev13}
\eeq
Taking the limit of this expression as $B_1 \rightarrow 0$, after doing the summation over
$n_1$, we get
\beqar
\Ree (i S_{\rm eff})  &=& - \int d^4x \, {E^2 \over 8 \pi^3} \sum_{n =1}^\infty {1\over n^2} e^{- m^2 (n \pi /E)}
\label{rev14}\\
&=& - \int d^4x\, {E^2 \over 48 \pi}, \hskip .2in {\rm as}~ m \rightarrow 0\nonumber
\eeqar
using $\sum_{n=1}^\infty (1/n^2) = \pi^2/6$ in the last line.
Again, this is in agreement with the standard Schwinger calculation.

The mass of the particle gives an exponential suppression of the pair creation rate, as is evident from
(\ref{rev9}) and (\ref{rev13}). To isolate and highlight the effect of the magnetic field,
it is useful to consider the massless case, i.e., the limit $m^2 \rightarrow 0$.
We can write the formulae for this as
\beq
\Ree ( i S_{\rm eff}) = - \int d^4x\,{E^2 \over 192 \pi} \times \left\{
\begin{matrix} f_0 ( B_1/E )\\
4\, f_{1/2} (B_1/E) \\
\end{matrix} \right.
\label{rev15}
\eeq
where $f_0$ applies to the spin zero case and $f_{1/2}$ to Dirac spinor.
These functions are given by
\beqar
f_0 (x) &=& {24\, x\over \pi} \sum_{n=1}^\infty {(-1)^{n+1} \over n} \, 
\sum_{n_1} e^{- n \pi (2 n_1 + 1) x}\nonumber\\
&=& {24\, x\over \pi} \sum_{n_1=0}^\infty 
\log \bigl( 1+ e^{-(2 n_1 + 1) \pi x}\bigr)\label{rev16a}
\eeqar
\beqar
f_{1/2} (x) &=& {6\,  x\over \pi} \sum_{n=1}^\infty {1\over n} 
e^{- m^2 (n \pi /E)} \sum_{n_1} \left(e^{- 2\pi n n_1 x}  + e^{- ( n_1 +1) 2 \pi n x}\right)
\nonumber\\
&=&{6\,  x\over \pi}  \left[ -\log \bigl( 1- e^{- m^2 \pi /E}\bigr) - 2\, \sum_{n_1 = 1}\log\bigl(
1- e^{-m^2 \pi /E}e^{ - 2\pi n_1 x}\bigr) \right]\label{rev16c}\\
&\approx&{6\,  x\over \pi}  \left[ -\log (m^2 \pi /E)- 2\, \sum_{n_1 = 1}\log\bigl(
1- e^{ - 2\pi n_1 x}\bigr) \right]
\label{rev16b}
\eeqar
The last line applies for $(m^2 \pi /E)\ll 1$. 

The motivation for the factorization of the decay rate in terms of these functions $f_0 (x)$,
$f_{1/2}(x)$ is that they
become $1$ as $x \rightarrow 0$, with
(\ref{rev15}) becoming identical to the standard Schwinger result.
These functions can thus be used to characterize the deviation from the case when
the background magnetic field is zero.
In (\ref{rev16c}), (\ref{rev16b}), we have also
separated off the contribution due to the zero mode for $f_{1/2}(x)$, namely,
for $n_1 = 0$ and also kept a nonzero value for the mass for this part.
This is because the zero mode contribution diverges at finite $B_1$ if the mass is zero.
This is an infrared divergence. Physically speaking, it is not sensible to consider a
uniform magnetic field over all of space. We must consider a finite volume, or, if
we wish to idealize a uniform magnetic field over a large volume as a constant value
over all of space, we should introduce an infrared cutoff. This is what is done
with the $m^2$-dependence term in (\ref{rev16c}) and the subsequent
simplification for small 

If we consider these functions at a fixed value of $E$ but vary $B_1$, it is easy to see
that $f_0(x)$ is always less than one. Thus the effect of the magnetic field
is to suppress the pair production rate.
This is straightforward to understand. The electric field has to create pairs which go into
various Landau levels, the most favorable would be the lowest Landau level
with the zero-point energy $B_1$. This energy cost suppresses the pair production
even if the mass is zero.
For the spin-$\half$ case, there is a zero mode, so there is no energy cost
for producing pairs which occupy this mode. Since the particles are fermions, there is
a limit given by the degeneracy proportional to the total area
of the $(x_1,x_2)$-subspace. So we get a divergent rate for pair production unless we cutoff
the area via an infrared cutoff.
Notice that, as the value of $B_1$ increases, all terms in the summation
in $f_{1/2}(x)$ get exponential suppression factors, except for
the zero mode part corresponding to
$ [ - (6  /\pi) \log(m^2 \pi /E)]\,x $. This leads to a linear increase of
$f_{1/2}$ with $B_1$ showing that there is enhancement of pair production.
The linear dependence in $B_1$ can be understood as due to the increase of degeneracy
as $B_1$ increases.

\subsection{Pair Creation on $T^2 \times {\mathbb R}^{1,1}$}

Before we start to examine the curvature effects, it is instructive to study if and how the toroidal compactification of directions transverse to the electric field influences the pair production rates.  

To compute the one-loop effective action on $T^2 \times {\mathbb R}^2$, we follow the same strategy as in the previous sections and consider uniform magnetic fields on $T^2$ and ${\mathbb R}^2$ denoted by $B_1$ and $B_2$, respectively. 
In the presence of uniform magnetic field $B_1$ on $T^2$, suitable boundary conditions have to be imposed on wave functions. A well-known choice is to implement periodic boundary conditions under magnetic translations \cite{Jain}, to which we confine our discussion in this subsection. A simple consequence of this type of boundary conditions is that, it leads to the Dirac quantization condition on the magnetic fields, that is, we have $B_1 = \frac{N}{2 \pi a^2}$ with $N \in \mathbb{Z}$, where $a$ stands for the compactification radius in each circular direction of $T^2$. 

The spectrum of the Laplace operator on $T^2 \times {\mathbb R}^2$ with and without the transverse magnetic field background is given as
\be
{\cal\mbox{Spec}}(-D^2) = 
\begin{dcases}
	\frac{1}{a^2} \left (p^2 + q^2 \right) + (2n + 1)B_2 \,, \quad & B_1 = 0  \,, \quad  p\,, q  \in \mathbb{Z} \,, n \in \mathbb{Z}_+ \,, \\
	(2n_1+1) \frac{N}{2 \pi a^2} + (2n_2 + 1)B_2  \,, \quad & N \neq 0 , \quad n_1 \,, n_2 \,  \in \mathbb{Z}_+ \,,
\end{dcases}
\label{SpecLapT}
\ee
From the first line of (\ref{SpecLapT}), we see that, in the absence of any magnetic flux penetrating $T^2$, there is a single zero mode specified by the quantum numbers $(p,q)=(0,0)$.
Once the magnetic field on $T^2$ is switched on though, the spectrum is formally the same as that of the flat case, except that $B_1$ is quantized as we have already noted. The corresponding density of 
states, for each state labeled by $(p,q)$ when $B_1 = 0$, is given by
\be
\rho_{T^2} =
\frac{1}{4 \pi^2 a^2} 
\ee
For the case with $N \neq 0$, we have the density of states as before,
\be
\rho_{T^2} = {B_1 \over 2 \pi} =
 \frac{N}{4\pi^2 a^2} \,, \quad \,N \neq 0 \, ,
\ee
Following the same steps as before, we find
\be
{\Ree}(i S_{\rm eff}) = - \frac{E^2}{16 \pi^3} \int_{T^2 \times {\mathbb R}^{1,1}} d^4 x \, \beta (\omega) \,,
\label{t2effa}
\ee
where, for $B_1 =0$,
\be
\beta {(B_1=0)}(\omega) := \frac{\omega}{\pi} \big (\log 2 + 4 \sum_{p=0}^{\infty} \log [1 + e^{ - \omega \left(p^2  + m^2 a^2 \right)}] + 4 \sum_{(p, q) > (0,0)}^{\infty} \log [1 + e^{ - \omega \left ( p^2 + q^2 + m^2 a^2 \right)}] \big) \,,
\label{betaT0}
\ee
and, for $B_1 = \frac{N}{2 \pi a^2}$,
\be
\beta(\omega) := \frac{\omega}{\pi} N \sum_{k = 0}^{\infty} \log [1 + e^{ - \frac{\omega}{\pi} ( N (k+ \frac{1}{2}) + m^2 a^2)}] \,, 
\label{betaT}
\ee
with $\omega := \frac{\pi}{E a^2}$.

\begin{figure}[!htb]
	\centering
	\includegraphics[width=0.65\linewidth,height=0.25\textheight]{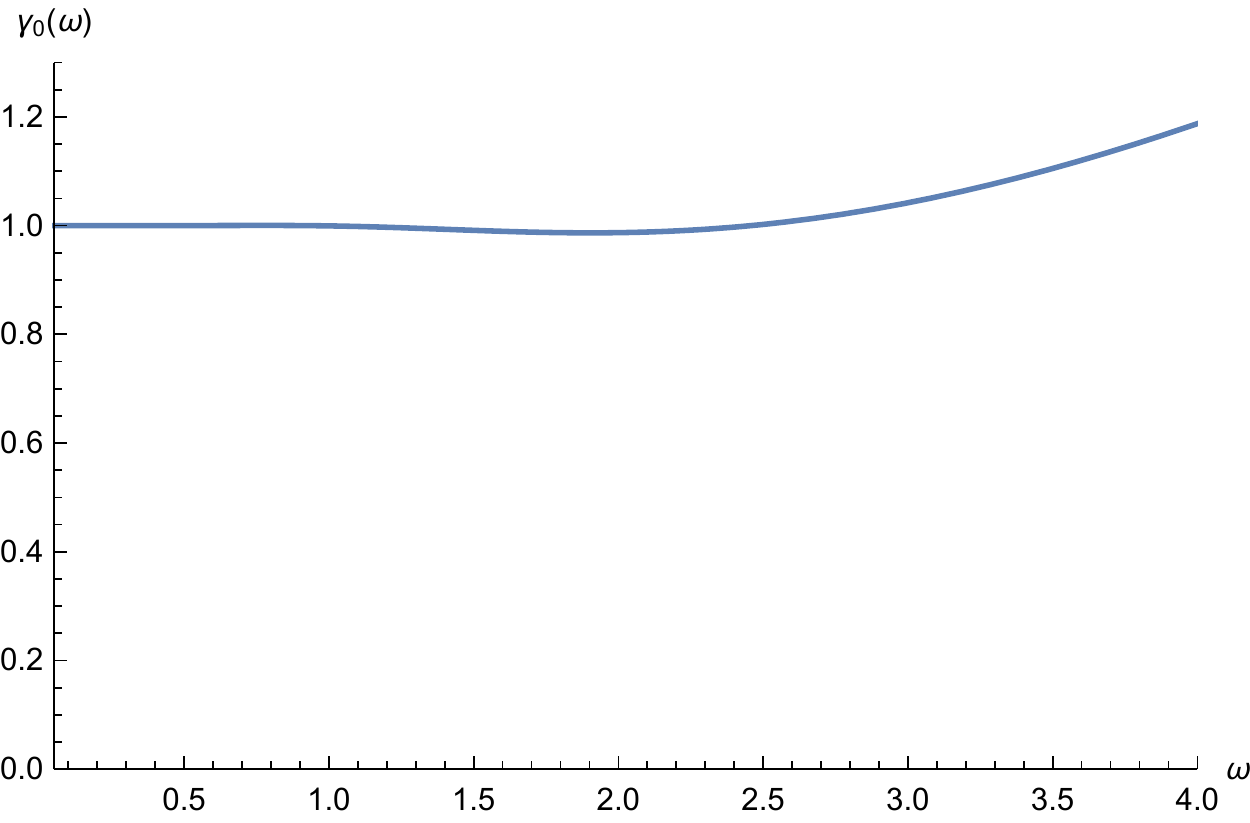}
	\caption{$\gamma(\omega)$ versus $\omega$ on $T^2 \times {\mathbb R}^{1,1}$}
	\label{T2M2_Scalar_Fig1}
\end{figure}

Let us examine the case with $B_1=0$ first. To probe the effect of compactification we take the ratio of $\beta_{(B_1=0)}(\omega)$ to the corresponding quantity $\frac{\pi^2}{12}$ for the ${\mathbb R}^{3,1}$ case, which is computed by the sum in (\ref{rev10}). We have
\beqa
\gamma(\omega) : &=& \frac{12}{\pi^2} \beta_{(B_1=0)}(\omega, m^2=0) \nn \\
&=& \frac{12}{\pi^3} \omega  \big (\log 2 + 4 \sum_{p=0}^{\infty} \log [1 + e^{ - \omega p^2}] + 4 \sum_{(p, q) > (0,0)}^{\infty} \log [1 + e^{ - \omega \left ( p^2 + q^2 \right)}] \big) \,.
\eeqa
We get a good estimate of $\gamma(\omega)$, by performing the sum over the discrete momenta $(p,q)$ up to $(p,q)_{max} =(1000,1000)$. 
This gives the profile presented in (Figure \ref{T2M2_Scalar_Fig1}).
Notice that the result is essentially the same as the Minkowski (${\mathbb R}^{3,1}$)
result
at small values of $\omega$. This is understandable since small $\omega$, at fixed
$E$, corresponds to
large $a$, and hence there should not be any significant
effect due to the compactification.
As $\omega$ increases, although, $T^2$ is flat, the change of the spectrum of the Laplacian 
due to compactification leads to an increase of the
 pair production amplitude on $T^2 \times {\mathbb R}^{1,1}$.

We observe (\ref{betaT}) with $m^2=0$ is the same as that is found on ${\mathbb R}^{3,1}$ in (\ref{rev15}) with $f_0(x)$ given in (\ref{rev16a}) with the replacement that $B_1 = \frac{N}{2 \pi a^2}$. Thus, there is essentially no change in the pair production effect on $T^2 \times {\mathbb R}^{1,1}$ compared to ${\mathbb R}^{3,1}$ as long as the transverse magnetic field is present, except that only quantized values of the magnetic field are admissible in the toroidally compactified setting.

In the absence of magnetic field, the Dirac operator on $T^2 \times {\mathbb R}^{2}$  squares to the Laplace operator on this space, since neither Zeeman-type nor curvature contributions are
 present in this case. Thus, the pair production effect is given by (\ref{t2effa}) for the spin up and spin down components. Finally, in the presence of the transverse magnetic field, the square of the Dirac operator has the same spectrum as the one obtained on ${\mathbb R}^{4}$ in (\ref{rev11a}), except for the quantized values of the magnetic field, i.e., $B_1 = \frac{N}{2 \pi a^2}$. Pair production effect, is therefore given by (\ref{rev15}), (\ref{rev16c}).
	
\section{Pair Creation on $S^2 \times {\mathbb R}^{1,1}$}

\subsection{Scalar Field, spin zero}

For the manifold $S^2 \times {\mathbb R}^{1,1}$, we start with considering uniform magnetic fields on $S^2$ 
and ${\mathbb R}^{2}$, which we label, as before, by $B_1$ and $B_2$, respectively. On the sphere, the uniform magnetic field $B_1$ is that of a magnetic monopole and therefore given by $B_1 = \frac{N}{2 a^2}$, with $N \in {\mathbb Z}$, due to the Dirac quantization condition. 
The Landau problem can be solved exactly using group theory, utilizing the fact that
$S^2 = SU(2)/U(1)$. The wave functions are the representation matrices
for $SU(2)$ of the form
$\bra{j, i} {\hat g} \ket{j, - {\half}N}$, for ${\hat g} \in SU(2)$, and $j = k + (N/2)$
\cite{KN}.
The spectrum of the gauged Laplacian, $-D^2$, is then readily obtained as 
\be
{\cal\mbox{Spec}}(-D^2) = 
\frac{1}{a^2} \left ( k(k+1) + N k + \frac{N}{2} \right) + (2 n_2 + 1)B_2 
\label{SpecLapS}
\ee
where $k, n_2$ take integer values from zero to infinity.
The density of the states is given by $\frac{B_2}{2\pi}$ on ${\mathbb R}^{2}$; on $S^2$ each Landau level has degeneracy $2k+1+N$ which is the dimension of the spin $j = k + \frac{N}{2}$ irreducible representation. Therefore for the density of states we have 
\be
\rho_{{\mathbb R}^{2}} = \frac{B_2}{2\pi} \,, \quad \rho_{S^2} = \frac{2k+1+N}{4 \pi a^2} \,.  
\label{Sp1}
\ee
The one-loop effective action is given by
\beqar
\Gamma &=&\Tr \,\log (-D^2 + m^2)\nonumber\\
&=&-\frac{1}{16 \pi^2 a^2} \int_{S^2} d\mu \int_{{\mathbb R}^{2}} d x_3 \, d x_4
\int \frac{ds}{s} \frac{B_2}{\sinh \, s B_2}\nonumber\\
&&\hskip .5in \times  \sum_{k=0}^{\infty} (2 k + 1+ N) e^{- s \left[ m^2 + \left( k(k+1) + N k + \frac{N}{2}\right)/a^2 \right]} \,. 
\label{EEA1S}
\eeqar
As before, we continue from $S^2 \times {\mathbb R}^{2}$ to Minkowski signature
for the ${\mathbb R}^{2}$-part using the Wick rotation
$B_2 \rightarrow -i E$ and $x_4 \rightarrow i x_0$. The real part of
$i S_{\rm eff}$ is then obtained as
\beq
\Ree (i S_{\rm eff})  =
\int d\mu dx_0 dx_3\,  {E\over 16 \pi^2 a^2}\, \sum_{n =1}^\infty
{(-1)^n \over n}  \, H(n\pi/E)
\label{EEA2S}
\eeq
where, we have defined
\be
H(s) = \sum_{k=0}^{\infty} (2 k + 1+ N)\, e^{- s \left[ m^2 +
	\left( k(k+1) + N k + \frac{N}{2}\right)/a^2 \right]}
\label{Sp2}
\ee
This can be rewritten as
\beqar
{\Ree}(i S_{ \rm eff}) 
&=& - \int d\mu dx_0 dx_3 \, {E^2 \over 16 \pi^3}\, \beta_0 (\omega)
\label{Sp3}\\
\beta_0 (\omega)&=& \omega \, \sum_{k =0}^\infty (2 k +1 +N) \sum_{n =1}^\infty {(-1)^{n+1}\over n}\,
e^{- n \omega \left[ m^2 a^2 + k (k+1) + N k + (N/2)\right]}\nonumber\\
&=& \omega \, \sum_{k =0}^\infty (2 k +1 +N)  \log [ 1 + e^{- \omega \left[ m^2 a^2 + k (k+1) + N k + (N/2)\right]}]\label{Sp4}
\eeqar
We defined $\omega := \frac{\pi}{E a^2}$ as a convenient dimensionless variable.\footnote{
As a check on this formula, notice that if we take the limit of $a^2 \rightarrow \infty$,
$N \rightarrow \infty$ keeping $B_1 = N/2 a^2$ fixed
in the summand and then carry out
the summation, we find that $\beta_0 (\omega) \rightarrow (\pi^2 /12) f_0 (x)$
with $f_0 (x)$ given in (\ref{rev16a}), say, for $m^2 =0$. In this way, we recover the flat space result with
a magnetic field.}
We can compare our result with the flat space result by
considering the latter over an area $4\pi a^2$. Writing $B_1 = N/ 2a^2$, the flat space formula
gives
\beqar
\beta^{\rm flat}_0 (\omega ,  m^2 = 0 ) &=&
\omega \,N \sum_{k =0}^\infty \sum_{n =1}^\infty {(-1)^{n+1}\over n}\,
e^{- n \omega \left[  N k + (N/2)\right]} \nonumber\\
&=& \omega N \, \sum_{k=0}^\infty \log [ 1 + e^{- \omega [N k + (N/2) ]}]
\label{Sp5}
\eeqar
The ratio of $\beta_0 (\omega, m^2 = 0)$ to $\beta^{\rm flat}_0 (\omega , m^2 = 0)$ 
is a good measure for
the effect of curvature for non-vanishing magnetic field background and we define this ratio as 
\beq
\gamma_{0}(\omega):= {\sum_{k =0}^\infty (2 k +1 +N)  \log [ 1 + e^{- \omega \left[ k (k+1) + N k + (N/2)\right]}] \over  N \sum_{k'=0}^\infty \log [ 1 + e^{- \omega [N k' + (N/2) ]}]} \,, \quad N \neq 0 \,,
\label{Sp6}
\eeq
while in the absence of the magnetic background, i.e. for $N=0$, we have
\be
\gamma_{0}(\omega) := \frac{12}{\pi^2} \omega \sum_{k = 0}^\infty (2k+1) \log [ 1 + e^{- \omega  k(k+1)}] \,.
\ee
\begin{figure}[!htb]\centering
	\begin{minipage}[t]{0.48\textwidth}
		\centering
		\includegraphics[width=1\textwidth]{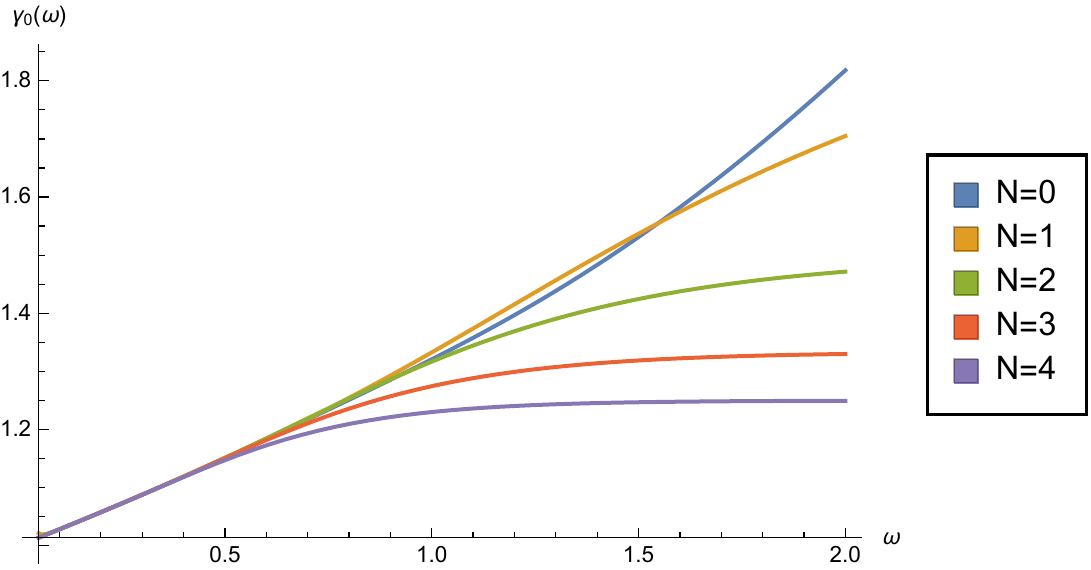}
		\caption{$\gamma_0(\omega)$ versus $\omega$.} 
		\label{gamma_01}
	\end{minipage}\hskip .1in
	\begin{minipage}[t]{0.48\textwidth}
		\centering	
		\includegraphics[width=1\textwidth]{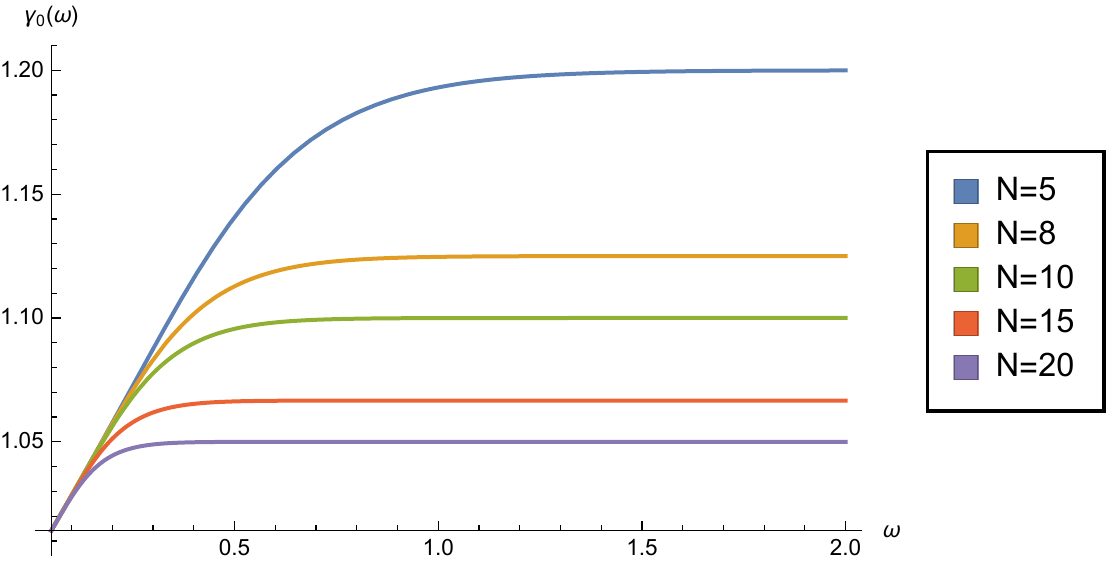}
		\caption{$\gamma_0(\omega)$ versus $\omega$.} 
		\label{gamma_02}
	\end{minipage}
\end{figure}

In Figs.\,\ref{gamma_01} and \ref{gamma_02}, we show the behavior of this ratio for several values
of $N$. {\sl Clearly, there is an enhancement effect due to the curvature}. 
This is basically due to the degeneracy factor $(2k +1 +N)$ on the sphere. In terms of
the magnetic field this is $(B/2\pi) + (2 k+1)/ 4 \pi a^2$ compared to
just $B/2\pi$ for the flat case. 
As $\omega$ becomes large, the $k = 0$ term dominates in both
the numerator and denominator in
(\ref{Sp6}), clearly showing that
the
$\gamma_0$'s saturate to a value of $\frac{N+1}{N}$.

\subsection{Dirac Field, spin $\half$}

We now turn to the case of the Dirac field on $S^2 \times {\mathbb R}^{2}$.
Again, the $S^2$-dependence of the
wave functions can be constructed in terms of the representation matrices
$\bra{j, m} {\hat g} \ket{j,m'}$ for $SU(2)$. The derivatives act as right translation
operators on ${\hat g}$. This has been used before for the solution of the
Landau problem for the scalars. For the Dirac case,
we have
\beq
\Gamma = - \Tr \log (i \gamma\cdot D +m ) =
{1\over 2}\int {d s\over s}  \Tr \left[ e^{-s (m^2 - (\gamma\cdot D)^2)}\right]
\label{Sp-Dir1}
\eeq
The square of the Dirac operator can be simplified as
\beq
- (\gamma\cdot D)^2 = - D_3^2 - D_4^2 - B_2 \left(\begin{matrix}
\sigma_3 &0\\ 0&-\sigma_3\\ \end{matrix} \right)
+ {1\over a^2} \left[( R_1^2 + R_2^2) - R_3  \left(\begin{matrix}
\sigma_3 &0\\ 0&\sigma_3\\ \end{matrix} \right)\right]
\label{Sp-Dir2}
\eeq
We have used the identification $D_a = i R_a/a$ for $a= 1, 2$. $R_3$, which arises from the
commutator of these operators, carries the information about the background magnetic
field on $S^2$ as well as the spin-curvature coupling.
The eigenvalues for $R_3$ are thus
$-{\half} - (N/2)$ for the first and third components of the spinor, and ${\half} - (N/2)$
for the other two components. Correspondingly, the $j$-value of the
representation must be $j = q+ (N+1)/2$ and $j = q+ (N-1)/2$,
where $q$  is a positive integer or zero. 
The eigenvalues and the density of states are then the following:
\beqar
( - (\gamma\cdot D)^2, \rho )
&=&\Bigl(  2 n_2 B_2 +  {1\over a^2} ( (q+1)^2 + N (q+1)),~  {2 (q+1)+ N \over 4 \pi a^2} \, {B_2 \over 2 \pi} \Bigr)\nonumber\\
&=&\Bigl( (2n_2 +2)B_2 + {1\over a^2} ( q^2 + N q), ~  {2 q+ N \over 4 \pi a^2} \, {B_2 \over 2 \pi}\Bigr)
\nonumber\\
&=& \Bigl( (2n_2 +2)B_2 + {1\over a^2} ( (q+1)^2 + N (q+1)),~  {2 (q+1)+ N \over 4 \pi a^2} \, {B_2 \over 2 \pi} \Bigr)\nonumber\\
&=&\Bigl(  2 n_2 B_2 + {1\over a^2} ( q^2 + N q), ~  {2 q+ N \over 4 \pi a^2} \, {B_2 \over 2 \pi}\Bigr)
\label{Sp-Dir3}
\eeqar
In the second and fourth of these equations, $q = 1, 2, $ etc. for $N = 0$,
while $q = 0, 1, 2$, etc. for $N\geq1$.
In the absence of a magnetic field, there is no zero mode on $S^2$. But for $N \neq 0$
there is a zero mode of degeneracy $N$ for the second and fourth components
of the Dirac spinor \cite{Dolan:2003bj}.
The effective action is then obtained as
\beqar
\Gamma &=& {1\over 16 \pi^2 a^2} \int d\mu dx_3 dx_4\int {d s\over s}
\, { B_2 \coth sB_2}\,
\sum_q \Bigl[ (2 q +N) e^{- s [m^2  + (q^2 + N q)/a^2 ]}\nonumber\\
&&\hskip 1.5in + (2 (q+1) +N ) e^{- s [m^2  + ((q+1)^2 + N (q+1) )/a^2 ]}
\Bigr]
\label{Sp-Dir4}
\eeqar
At large values of $a^2$ with fixed $N$, it is possible to treat $p= q/a$ as a continuous variable
and
convert the sum over $q$ to an integration. It is then easy to check that this expression agrees
with what was obtained for flat space. We can extract the decay rate due to pair production
as before by evaluating the contribution to integral over $s$
 from the poles from $(\sin sE)^{-1}$ after continuation to Minkowksi space.
 The result is then
 \beqar
 \Ree (i S_{\rm eff}) &\!\! =&\!\! - \int d\mu dx_0 dx_3 \,
 {E^2\over 8 \pi^3} \, \beta_{1/2}(\omega )\nonumber\\
 \beta_{1/2}(\omega) &\!\! =&\!\!  \omega \sum_{n=1}^\infty
 {1\over n} e^{- n \omega (m^2 a^2)} \left[
 {N\over 2} + \sum_{q=1}^\infty (2 q +N) e^{- n \omega (q^2 + N q)}\right]\nonumber\\
 &\!\! =&\!\!  - \omega \left[{N\over 2}  \log \bigl( 1- e^{-\omega m^2 a^2}\bigr)
 + \sum_{q=1}^\infty (2 q+N) \log \bigl( 1- e^{-\omega ( q^2 + N q + m^2 a^2)}\bigr)
 \right]
 \label{Sp-Dir5}
 \eeqar
 where $\omega = (\pi /Ea^2)$.
 In this expression for $\beta_{1/2}(\omega)$, the first term
 $- \omega \log ( 1- e^{- \omega m^2 a^2})$ is the contribution of
 $q =0$ in the first and last set of eigenvalues in (\ref{Sp-Dir3}); i.e., it is due to
 the zero mode of $(\gamma\cdot D)^2$ on $S^2$. Notice that this term diverges
 if we take $m^2 \rightarrow 0$, very similar to what we found
for the case of flat space.
For the second term with the summation over $q$, we can set $m^2 = 0$ if we want to consider
the massless case, without loss of convergence. 

\begin{figure}[!htb]\centering
		\centering
		\scalebox{.6}{\includegraphics[width=1\textwidth]{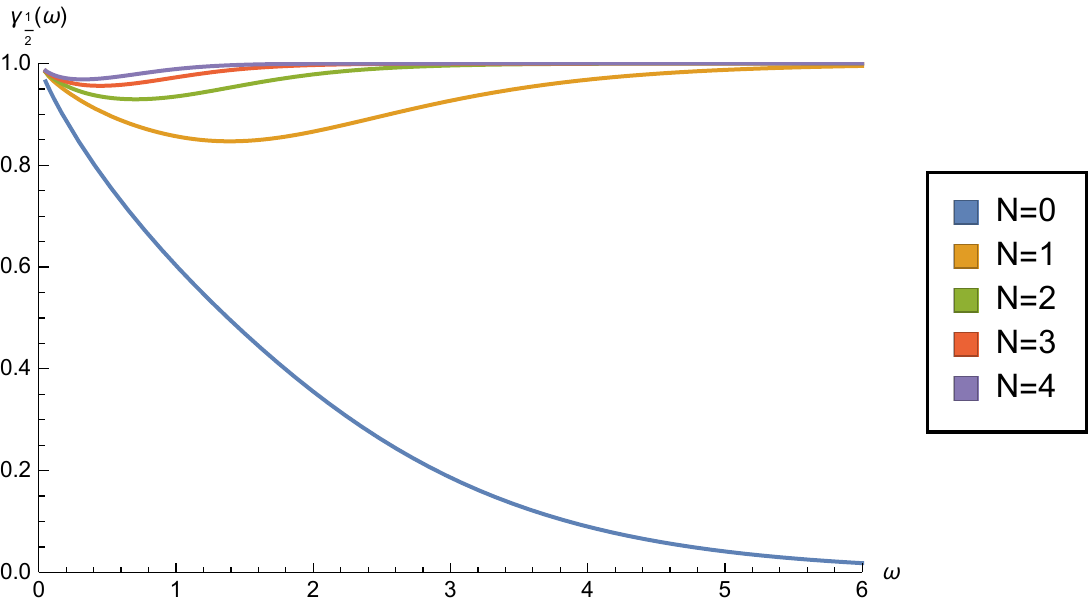}}
		\caption{$\gamma_{1/2}$ versus $\omega$.} 
		\label{gamma_half1}
\end{figure}

The limit of $\beta_{1/2}(\omega)$ for flat space with a magnetic field,
of flux $N$ over area $4\pi a^2$, is given by
\beq
\beta^{\rm flat}_{1/2}(\omega)
= - \omega \left[{N\over 2}  \log \bigl( 1- e^{-\omega m^2 a^2}\bigr)
+ N  \sum_{q=1}^\infty \log \bigl( 1- e^{-\omega ( N q + m^2 a^2)}\bigr)
 \right]
 \label{Sp-Dir6}
 \eeq
We can now define the ratio
\beq
\gamma_{1/2}(\omega ) = { {N}  \log \bigl( 1- e^{-\omega m^2 a^2}\bigr)
 + 2 \sum_{q=1}^\infty (2 q+N) \log \bigl( 1- e^{-\omega ( q^2 + N q + m^2 a^2)}\bigr) \over
 {N}  \log \bigl( 1- e^{-\omega m^2 a^2}\bigr)
+ 2 N  \sum_{q'=1}^\infty \log \bigl( 1- e^{-\omega ( N q' + m^2 a^2)}\bigr)}
\label{Sp-Dir7}
\eeq
We have already seen that there was an enhancement of pair production due to
the zero mode in flat space. $\gamma_{1/2}(\omega) $ gives a measure of the effect of
curvature. We show the behavior of this function for several values of $N$ in
Fig.\,\ref{gamma_half1}. For these graphs, we have taken
$m^2 a^2 = \frac{1}{2}$ as the cut-off value.
The summations were carried out to $(q_{max}, q'_{max}) = (100,100)$.
The case of zero magnetic field is special since there
are no zero modes. Therefore, in order to assess the effect of curvature, we can compare $\beta_{1/2}(\omega, N=0, m=0)$ with $\beta_{1/2}^{\rm flat}(\omega, m=0) = \sum_{n=1}^\infty \frac{1}{n^2} = \frac{\pi^2}{6}$ and examine the profile of 
\beq
\gamma_{1/2}(\omega, N=0) = - \frac{\beta_{1/2}(\omega, N=0)}{({\pi^2}/{6})} 
= - \frac{12}{\pi^2} \omega \sum_{q=1}^\infty q \log \bigl( 1- e^{-\omega q^2}\bigr) \,.
\eeq
The result is also shown in Fig.\,\ref{gamma_half1}.
From these plots of $\gamma_{1/2}(\omega)$, we can infer that:
\begin{itemize}
\item In the absence of any transverse magnetic field, the pair production effect is significantly diminished compared to the flat case.
\item When the magnetic field is present, pair production effect still remains less than that on the flat space at any given magnetic field.
\end{itemize}
These can be seen as a consequence of the curvature of $S^2$.  
For small values of $\omega$, the $q^2$ dependence of the
eigenvalues (which is due to the curvature effects) ensures that
the numerator in (\ref{Sp-Dir7}) remains smaller than the denominator, giving
$\gamma_{1/2}(\omega) < 1$.
We may further observe that, at any fixed magnetic field, the effect tends to converge to the flat space result with increasing $\omega$.
This can be attributed to the fact that the contribution from the zero modes dominates
as $\omega$ increases. At higher values of the magnetic charge, there are more zero modes and the restoring effect of the zero modes becomes stronger, narrowing the interval of $\omega$ in which $\gamma_{1/2}(\omega) <1$.
While there is suppression due to the positive curvature, it is worth noting
that the overall rate with a nonzero
background magnetic field is still higher than that for zero magnetic field.
This is clear from the limit of $\gamma_{1/2}(\omega)$ approaching the flat space value
for large $\omega$.


\section{Pair Creation on $H^2 \times {\mathbb R}^{1,1}$}

\subsection{Scalar Field, spin zero}

On the Euclidean space $H^2 \times {\mathbb R}^{2}$, again we label the directions on $H^2$ as $1, 2$ and directions on ${\mathbb R}^{2}$ as $3, 4$, without reference to any particular coordinate system. With a magnetic field $B_1 = \frac{b}{a^2}$ on $H^2$ and a magnetic field $B_2$ on ${\mathbb R}^{2}$, spectrum of the gauged Laplacian, $-D^2$, is
\be
{\cal\mbox{Spec}}(-D^2) = 
\begin{dcases}
	\frac{1}{a^2} (\lambda^2 + \frac{1}{4} + b^2) + (2n + 1)B_2 \,, \quad  \,& 0 \leq \lambda <\infty \\
	\frac{1}{a^2} \left (-(b-k-\frac{1}{2})^2 + \frac{1}{4} + b^2 \right) + (2n + 1)B_2, \! \quad \, & 0 \leq k \leq [b -\half] \,\, \mbox{and} \,\, k \in \mathbb{Z} \, .
\end{dcases}	
\label{SpecLapH}
\ee
The spectrum of the gauged Laplacian on $H^2$ is composed of a discrete and a continuous part. This problem is worked out in detail in the literature \cite{{Comtet:1984mm},{Comtet:1986ki}}, 
and we provide a brief review in the Appendix in order to keep the paper self-contained. The continuous part of the spectrum is labeled by the spectral parameter $\lambda$, while the discrete spectrum is labeled by the index $k$, which we will call the Landau level index on $H^2$. Contrary to the flat case, the LL index on $H^2$ does not extend to infinity, but is truncated by the largest integer less 
than $b -\frac{1}{2}$, i.e. by $[b-\frac{1}{2}]$. Thus discrete states exist only if $b > \frac{1}{2}$. In (\ref{SpecLapH}), $a$ stands for the radius of curvature of $H^2$. 

The density of states of the continuous as well as the discrete part of the spectrum of the gauged Laplacian on $H^2$ can be determined from the representation theory of the group 
$SL(2, {\mathbb R})$, which is the universal covering group of the isometry group $SO(2,1) \simeq SU(1,1)$ of $H^2$, as discussed in the Appendix. These are given as
\beqa
\rho_{b}(\lambda) &=& \frac{1}{2\pi a^2}\frac{ \lambda \, \sinh 2 \pi \lambda}{\cosh 2\pi \lambda +  \cos 2 \pi b} \,, \quad b \neq {\mathbb Z} + \frac{1}{2} \,, \\
\rho_{b}(k) &=& \frac{1}{2\pi a^2}\left(b -k-\frac{1}{2} \right) \,, \quad b > \frac{1}{2} \,,
\quad 0\leq k \leq [b - \half ],
\label{dosH2}
\eeqa
In the $b \rightarrow 0$ limit, the spectrum of the Laplacian on $H^2$ is $\frac{1}{a^2} (\lambda^2 + \frac{1}{4})$ and the density of states becomes $\rho_{0}(\lambda) = (\lambda \tanh \pi \lambda)/2 \pi a^2$. Apart from the $\frac{1}{2 \pi a^2}$ factor, this is the Plancherel measure for the harmonic functions over $H^2$, while $\rho_{b}(\lambda)$ may be understood as the Plancherel measure for the sections of the $U(1)$-bundle on $H^2$ with curvature $b$. 
Unlike the case of $S^2$, there is no Dirac quantization
condition for the values of the background magnetic field $b$; generally,
 $b \in {\mathbb R}$.

\begin{figure}[!htb]\centering
	\begin{minipage}[t]{0.48\textwidth}
		\centering
		\includegraphics[width=1\textwidth]{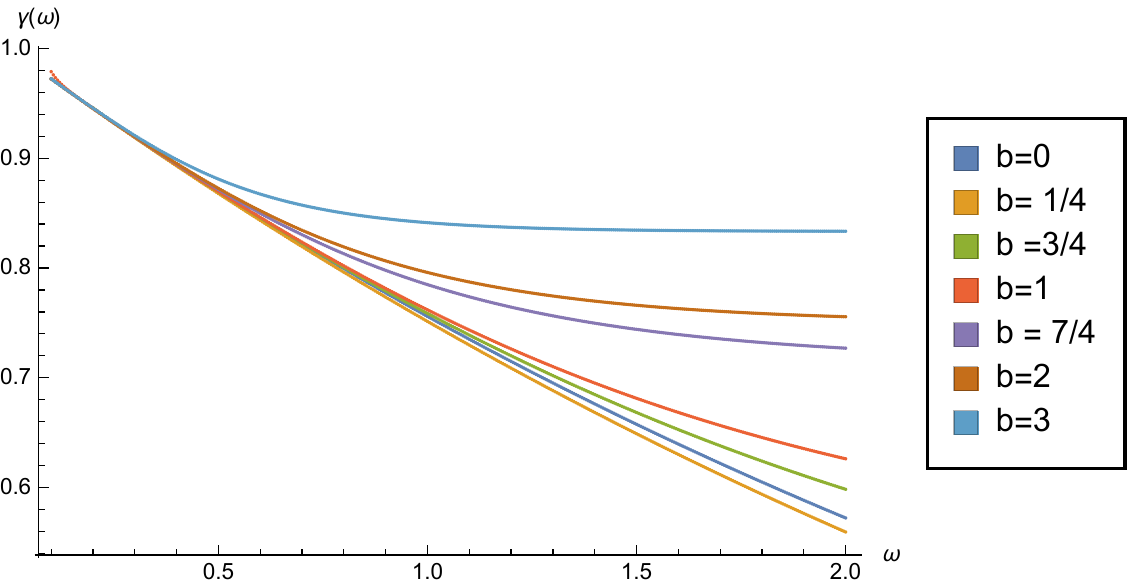}
		\caption{$\gamma_0(\omega)$ versus $\omega$.} 
		\label{H^2scalar1}
	\end{minipage} \hskip .1in
	\begin{minipage}[t]{0.48\textwidth}
		\centering	
		\includegraphics[width=1\textwidth]{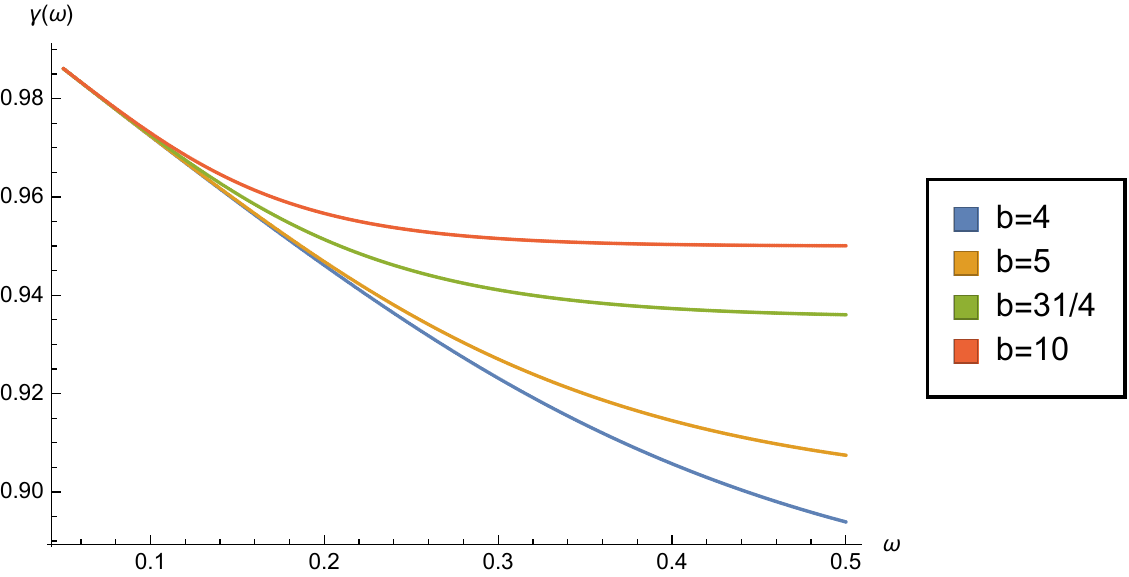}
		\caption{$\gamma_0(\omega)$ versus $\omega$.} 
		\label{H^2scalar2}
	\end{minipage}
\end{figure}

We are now in a position to take up the calculation of the one-loop effective action. This is given by
\beq
\Gamma = \mbox{Tr} \log (-D^2 + m^2) =
 - \mbox{Tr} \int \frac{ds}{s} e^{-s(-D^2+m^2)} \,.
\eeq
Given the spectrum of $-D^2$ and the density of states, this can be worked out as
\begin{multline}
\Gamma = -\frac{B_2}{8 \pi^2 a^2} \int_{H^2} d\mu \int_{R^2} d x_3 \, d x_4
\int \frac{ds}{s} \frac{e^{-s m^2}}{\sinh \, s B_2} \Bigg [\int_0^\infty d \lambda \, \frac{ \lambda \, \sinh 2 \pi \lambda}{\cosh 2\pi \lambda +  \cos 2 \pi b}
e^{-\frac{s}{a^2}(\lambda^2 + \frac{1}{4} + b^2)} \\
+ \sum_{k=0}^{\lbrack b-1/2 \rbrack} (b-k-\frac{1}{2}) e^{-\frac{s}{a^2} \left (-(b-k-\frac{1}{2})^2 + \frac{1}{4} + b^2 \right)} \,
\Bigg]\,,
\label{EEA1}
\end{multline}
where $d \mu$ stands for the volume form on $H^2$. We can write $\Gamma$ by introducing the short-hand notations $K_C(s)$ and $K_D(s)$ for the integral and the sum in the square-bracketed expression in (\ref{EEA1}) as
\be
\Gamma = -\frac{B_2}{8 \pi^2 a^2} \int_{H^2} d\mu \int_{R^2} d x_3 \, d x_4
\int \frac{ds}{s} \frac{e^{-s m^2}}{\sinh \, s B_2} (K_C(s) + K_D(s)) \,.
\label{EEA2}
\ee 
Continuing $H^2 \times {\mathbb R}^{2}$ to $H^2 \times {\mathbb R}^{1,1}$ by the
Wick rotation $B_2 \rightarrow -i E$ and $x_4 \rightarrow i x_0$, we may write the effective action as
\begin{multline}
i S_{\rm eff} = - \Gamma \Big|_{B_2 \rightarrow -i E \,, x_4 \rightarrow i x_0} =
\frac{i E}{8 \pi^2 a^2} \int_{H^2} d\mu \int_{M^2} d x_0 \, d x_3
\int \frac{ds}{s} \frac{e^{-s m^2}}{\sin  s E}\, (K_C(s) + K_D(s)) \\
- (\mbox{renormalization corrections}) \,.
\label{EEA3}
\end{multline} 
As in other cases, the imaginary part of the contribution to the integral over $s$ can be obtained
from the residues at the poles of $\sin sE$. This leads to the result
\be
{\Ree}(i S_{\rm eff}) =
\frac{E}{8 \pi^2 a^2} \int_{H^2} d\mu \int_{{\mathbb R}^{1,1}} d x_0 \, d x_3\, \sum_{r=1}^\infty \frac{(-1)^r}{r} \bigl[ K_C(r \pi/E)+K_D(r \pi/E) \bigr] \,.
\label{EEA3}
\ee 
Carrying out the summation  over $r \in {\mathbb Z}_+$ this expression can be simplified to
\begin{multline}
{\Ree}(i S_{\rm eff}) =
- \frac{E}{8 \pi^2 a^2 } \int {d\mu} d x_0 \, d x_3\,
\Bigg [\int_0^\infty \!\!\!d \lambda \, \frac{ \lambda \, \sinh 2 \pi \lambda}{\cosh 2\pi \lambda +  \cos 2 \pi b} 
\log [1 + e^{-\frac{\pi}{E a^2}(\lambda^2 + \frac{1}{4} + b^2 + m^2 a^2)}] \\
+ \sum_{k=0}^{\lbrack b-1/2 \rbrack} (b-k-\frac{1}{2}) \log[1 + e^{-\frac{\pi}{E a^2} \left (-(b-k-\frac{1}{2})^2 + \frac{1}{4} + b^2 + m^2 a^2 \right)} \,
\Bigg]\,.
\end{multline}
As before, we introduce the
dimensionless variable $\omega := \frac{\pi}{E a^2}$ and write ${\Ree}(i S_{\rm eff})$ as 
\be
{\Ree}(i S_{\rm eff}) =
- \frac{E^2}{16 \pi^3} \int_{H^2} d\mu \int_{M^2} d x_0 \, d x_3\, \beta_0 (\omega)
\label{SEFFH2M2}
\ee
where $\beta_0(\omega) := \beta_C(\omega) + \beta_D(\omega)$ with
\beqar
\beta_C(\omega) &=&
2 \omega \int_0^\infty d \lambda \, \frac{ \lambda \, \sinh 2 \pi \lambda}{\cosh 2\pi \lambda +  \cos 2 \pi b} 
\log [1 + e^{-\omega(\lambda^2 + \frac{1}{4} + b^2 + m^2 a^2)}] \nonumber\\
\beta_D(\omega) &=& 2 \omega \sum_{k=0}^{\lbrack b-1/2 \rbrack} (b-k-\frac{1}{2}) \log[1 + e^{-\omega \left (-(b-k-\frac{1}{2})^2 + \frac{1}{4} + b^2 + m^2 a^2 \right)}]  
\label{betaH^2}
\eeqar
The flat limit of $H^2 \rightarrow {\mathbb R}^2$ with
$a^2 \rightarrow \infty$, and keeping a nonzero uniform perpendicular magnetic field $B_1$ 
on ${\mathbb R}^2$ is achieved by taking $b \rightarrow \infty$ such that $\frac{b}{a^2} \rightarrow B_1$. In this limit $\beta_C(\omega)$ gives no contribution at all as there is no continuous spectrum of energies in this limit, while the number of discrete states extends to infinity.
Thus retaining only the $b$-dependent terms in the energy spectrum and extending the sum over $k$ to infinity, we can write
\be
\beta^{\rm flat}_0 (\omega , m^2 = 0 ) = 2 \omega b \, \sum_{k=0}^{k_{max} \rightarrow \infty} \log [ 1 + e^{- \omega (2 b k + b)}] \,.
\label{H2Betalimit}
\ee
where we have also set $m^2=0$.
Notice that (\ref{H2Betalimit}) has the same form as (\ref{Sp5}) where $N$ is
replaced with $ 2 b$. Proceeding in the same manner as in the previous section, we take the ratio of these quantities and introduce
\be
\gamma_{0}(\omega) = \frac{\beta_0(\omega, m^2=0)}{\beta^{\rm flat}_0 (\omega ,  m^2 = 0 )} \,,
\ee
Profiles $\gamma_{0}(\omega)$ at several different values of the magnetic field $b$
 can
be obtained by evaluating the integral over $\lambda$ numerically. 
These are presented in Figs.\,\ref{H^2scalar1} and \ref{H^2scalar2}.

It is useful to consider the corresponding quantities for $b=0$ separately. Since, in this case, the Laplacian has only the continuous spectrum, we have
\be
\beta_0(\omega, b =0) = 2 \omega \int_0^\infty d \lambda \, \lambda \, \tanh  \pi \lambda \log [1 + e^{-\omega(\lambda^2 + \frac{1}{4} + m^2 a^2)}]
\ee
while from equation (\ref{rev16a}) or (\ref{Sp5}), we have already $\beta_0^{\rm flat}(\omega)|_{m^2 \rightarrow 0} = \frac{\pi^2}{12}$. Therefore, curvature effects can be probed using
\beqa
\gamma_{0}(\omega, b =0) &:=& \frac{\beta_0(\omega, m^2 \rightarrow 0)}{\beta_0^{\rm flat}(\omega, m^2 \rightarrow 0)} \,, \\
&=& \frac{24}{\pi^2} \omega \int_0^\infty d \lambda \, \lambda \, (\tanh  \pi \lambda )\, \log [1 + e^{-\omega(\lambda^2 + \frac{1}{4} + m^2 a^2)}] \,.
\eeqa
Profile of this function is also included in Fig.\,\ref{H^2scalar1}.

From the profile of $\gamma_0(\omega)$'s, we see that rate of pair production on $H^2 \times \mathbb R^{1,1}$ is always less than what it is in Minkowski space
${\mathbb R}^{3,1}$. This result is in contrast to the 
enhancement effect on $S^2 \times {\mathbb R}^{1,1}$ and, generally speaking, may be attributed to the constant negative curvature of $H^2$. Profiles of $\gamma_0(\omega)$'s for $H^2 \times {\mathbb R}^{1,1}$ show that the rate of decrease of $\gamma_0(\omega)$'s from the starting value of
one becomes less with the increasing value of the
$b (> \frac{1}{2})$-field, i.e., the pair production becomes relatively larger with the increasing $b$-field.
This is in contrast to what we observe on $S^2 \times {\mathbb R}^{1,1}$ and, seemingly,
counterintuitive to our general expectation of less pair production with increasing magnetic field, based on the fact that it is energetically costlier for particles to occupy LLs. (Recall
that even the LLL has energy $\sim b/a^2$.) Nevertheless, there is a simple way to see the underlying reason for this behavior of $\gamma_0(\omega)$'s.
At low values of the $b$-field ($>\frac{1}{2}$) on $H^2$ there are very few LLs and almost all available states are continuous energy levels similar to the case of the flat space with no magnetic field.
The degeneracy of the continuous states ($\sim d \lambda \, \lambda \tanh(\pi \lambda)$)
is less than what is obtained for continuous states
in flat space ($\sim dk\,k$); also the eigenvalues start at nonzero values ($\geq {1\over 4}$).
These two factors together lead to
a decrease in the pair production effect.
With increasing $b$-field, however, there are more and more LLs on $H^2$.
Although it is still energetically costly for the particles to fill them, the Landau levels have less energy compared to the flat case ($ \frac{1}{a^2}(-k(k+1) + 2bk +b) \leq  2 B_1 k +B_1$) and, in addition, it is less costly than filling the continuum energy levels, whose zero point energy is $(b^2 + \frac{1}{4})/a^2$. Thus, produced particles tend to fill these states, alleviating to an extent the sharper decrease in the pair production that happens in the absence of the transverse $b$-field.
The effect remains diminished compared to the flat case, but the deviation becomes 
less at larger values of $b$. 

The situation for $b <\frac{1}{2}$ is special because the advantage of discrete states
does not come in until $b$ exceeds $\half$.
There are no discrete energy states for $b < \half$, and it becomes harder for particles to fill in the continuous states due to the increasing energy cost, which causes a further decrease in the effect.
This accounts for the
lower rates for nonzero $b < \half$, compared to $b = 0$,
 as can be seen from the plot of the case with $b = \frac{1}{4}$. 
 
\begin{figure}[!htb]
	\centering
	\scalebox{.8}{\includegraphics[width=0.6\linewidth,height=0.25\textheight]{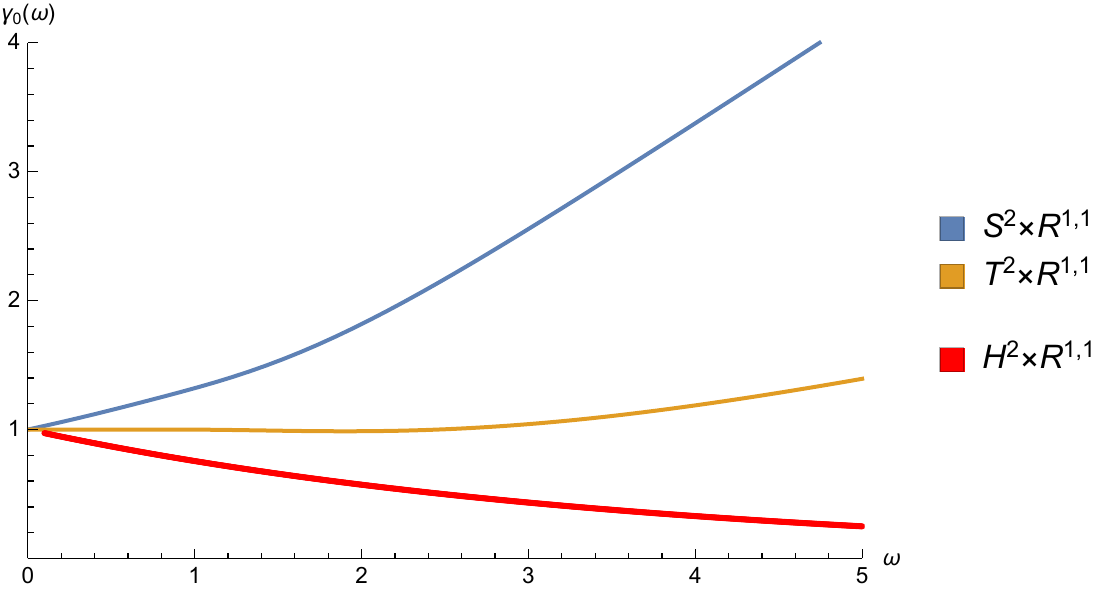}}
	\caption{Comparison of $\gamma_{0}$ at zero magnetic field for the torus, sphere and hyperboloid}
	\label{gamma_0N=0}
\end{figure}

 Finally we give a comparison of $\gamma_0 (\omega)$ for the three cases of
 torus, sphere and the hyperboloid for the case of zero magnetic
 field in Fig.\,\ref{gamma_0N=0}.

\subsection{Dirac Field, spin-$\half$}

The computation of the pair production amplitude for spin $\frac{1}{2}$ particles on $H^2 \times {\mathbb R}^{1,1}$ can be done, starting once again, with the Euclidean space $H^2\times {\mathbb R}^2$. 
The square of the gauged Dirac operator
now reads
\beq
- (\gamma\cdot D)^2 = - D_3^2 - D_4^2 - B_2 \left(\begin{matrix}
\sigma_3 &0\\ 0&-\sigma_3\\ \end{matrix} \right)
+ {1\over a^2} \left[( R_1^2 + R_2^2) + R_3  \left(\begin{matrix}
\sigma_3 &0\\ 0&\sigma_3\\ \end{matrix} \right)\right]
\label{H2-Dir}
\eeq
The spectrum of the square of this operator can be written in two parts. The continuous part
is given by
\be
{\cal\mbox{Spec}}_C(-\slashed{D}^2) = \frac{1}{a^2} (\lambda^2 + b^2) +
\begin{dcases}
	&2 n B_2 \,, \\
	&(2n+2) B_2
\end{dcases}	
\,, \quad \mbox{for} \,\, 0 \leq \lambda <\infty
\label{SpecDirac2HC}
\ee
The discrete part of the spectrum is given as
\be
{\cal\mbox{Spec}}_D(-\slashed{D}^2) = 
\begin{dcases}
	\frac{1}{a^2} \left (-(k+1-b)^2 + b^2 \right) + 2 n B_2 \,, \quad &\mbox{for} \,\, 0 \leq k \leq [b-1] \\
\frac{1}{a^2} \left (-(k-b)^2 + b^2 \right) + (2n + 2) B_2 \quad &\mbox{for} \,\, 0 \leq k \leq [b]  \\
	\frac{1}{a^2} \left (-(k+1-b)^2 + b^2 \right) + (2 n+2) B_2 \,, \quad &\mbox{for} \,\, 0 \leq k \leq [b-1]  \\
	\frac{1}{a^2} \left (-(k-b)^2 + b^2 \right) + 2 n B_2 \,, \quad &\mbox{for} \,\, 0 \leq k \leq [b]  \\
\end{dcases}	
\label{SpecDirac2HD}
\ee
The corresponding densities of states (for the $H^2$ part of the spectrum) are given by
\be
\rho^{(1/2)}_{b}(\lambda) = \frac{1}{2\pi a^2}\frac{ \lambda \, \sinh 2 \pi \lambda}{\cosh 2\pi \lambda - \cos 2 \pi b} \,, \quad \quad
\rho^{(1/2)}_{b}(k) \equiv \left(\frac{1}{2\pi a^2}\left(b-k-1\right) \,, \, \frac{1}{2\pi a^2}\left(b-k \right)\right) \,.
\label{dosspinorH2}
\ee
The first entry for $\rho^{(1/2)}_{b}(k) $ applies for the first and third component of the spinor and
the second entry for the second and fourth components.
When the magnetic field $b$ on $H^2$ is switched off, it should be clear that the discrete part of the spectrum of the Dirac operator on $H^2$ goes away and only the continuous part remains, whose eigenvalues are simply given as $\frac{\lambda^2}{a^2}$ and the density of states become $\rho^{(1/2)}(\lambda) = (\lambda \coth \pi \lambda)/2 \pi a^2$.
The effective action is given as
\beq
\Gamma = - \frac{1}{2} \, \mbox{Tr} \, \log (-\slashed{D}^2 + m^2) 
= \frac{1}{2} \, \mbox{Tr} \int \frac{ds}{s} e^{-s(\slashed{D}^2+m^2)} \,,
\eeq
Substituting from (\ref{SpecDirac2HC}),(\ref{SpecDirac2HD}) and (\ref{dosspinorH2}) this takes the form
\begin{multline}
\Gamma = \frac{B_2}{4 \pi^2 a^2} \int_{H^2} d\mu \int_{{\mathbb R}^2} d x_3 \, d x_4
\int \frac{ds}{s}  \coth\, s B_2 \, e^{-s m^2} \Bigg [\int_0^\infty d \lambda \, \frac{ \lambda \, \sinh 2 \pi \lambda}{\cosh 2\pi \lambda -  \cos 2 \pi b}
e^{-\frac{s}{a^2}(\lambda^2 + b^2)} \\
+ \frac{1}{2} b + \sum_{k=1}^{\lbrack b \rbrack} (b-k) \, e^{-\frac{s}{a^2} \left ( -(k-b)^2 + b^2\right)} \,
\Bigg]\,.
\label{EEAHSP1}
\end{multline}
In this expression, we have explicitly separated out the contribution coming from the zero modes of the Dirac operator on $H^2$, which are given by the $k=0$ terms in the second and fourth lines of (\ref{SpecDirac2HD}).

Introducing the short-hand notations, $K^{(1/2)}_C(s)$, $K^{(1/2)}_0(s)$ and $K^{(1/2)}_C(s)$ for the continuous part with the integral over $\lambda$, the zero-mode contribution and the discrete sum, respectively,
 in the square-bracketed expression in (\ref{EEAHSP1}), we can express $\Gamma$ as
\be
\Gamma = \frac{B_2}{4 \pi^2 a^2} \int_{H^2} d\mu \int_{R^2} d x_3 \, d x_4
\int \frac{ds}{s} e^{-s m^2} \, \coth \, s B_2 \, \left(K^{(1/2)}_C(s) + K^{(1/2)}_0(s) + K^{(1/2)}_D(s) \right) \,.
\label{EEASP2}
\ee 
Once again, we continue $(\Gamma )$ from
$H^2 \times {\mathbb R}^2$ to $H^2 \times {\mathbb R}^{1,1}$ by the
Wick rotation $B_2 \rightarrow -i E$ and $x_4 \rightarrow i x_0$, and obtain
\beqar
i S_{\rm eff}
&=& 
-\frac{i E}{4 \pi^2 a^2} \int_{H^2} d\mu \int_{{\mathbb R}^{1,1}} d x_0 \, d x_3
\int \frac{ds}{s} e^{-s m^2} \cot s E\,
\, \left ( K^{(1/2)}_C(s)  + K^{(1/2)}_0(s) + K^{(1/2)}_D(s) \right) \nonumber\\
&& \hskip 2in - (\mbox{renormalization corrections}) \,.
\label{EEAHSP3}
\eeqar
Evaluating the contribution from the residues of the poles of $\cot s E$
we find the real part of $i S_{\rm eff}$ to be 
\be
{\Ree}(i S_{\rm eff}) =
-\frac{E}{4 \pi^2 a^2} \int d\mu d x_0 \, d x_3\, \sum_{r=1}^\infty \frac{1}{r} \bigl[
K^{(1/2)}_C(r \pi /E) + K^{(1/2)}_0(r \pi /E) + K^{(1/2)}_D(r \pi /E) \bigr] \,.
\label{EEASP4}
\ee 
Using the expressions for the $K$'s from
(\ref{EEAHSP1}), we can carry out the summation over $r$. This leads to
\begin{multline}
{\Ree}(i S_{\rm eff}) =
\frac{E}{4 \pi^2 a^2 } \int  {d\mu}d x_0 \, d x_3\,
\Bigg [\int_0^\infty d \lambda \, \frac{ \lambda \, \sinh 2 \pi \lambda}{\cosh 2\pi \lambda - \cos 2 \pi b} 
 \log [1 - e^{-\frac{\pi}{E a^2}(\lambda^2 + b^2 + m^2 a^2)}] \\
 + \frac{1}{2} b \log [1- e^{-\frac{\pi}{E a^2}}] 
+ \sum_{k=1}^{\lbrack b \rbrack} (b-k) \log[1 - e^{-\frac{\pi}{E a^2} \left ( -(k-b)^2 + b^2 + m^2 a^2 \right)} \,
\Bigg]\,.
\end{multline}
Using $\omega = \frac{\pi}{E a^2}$, we can express this in a form similar to what we had for the
scalar case as
\be
{\Ree}(i S_{\rm eff}) =
- \frac{E^2}{8 \pi^3} \int_{H^2} d\mu \int_{M^2} d x_0 \, d x_3\, \beta_{1/2} (\omega)
\label{SEFFH2M2SP}
\ee
where
$\beta_{1/2}(\omega) := \beta_{1/2\,,C} (\omega) + \beta_{1/2\,,0}(\omega) + \beta_{1/2\,,D}(\omega)$
with
\beqa
\beta_{1/2\,,C} (\omega) &:=&
- 2 \omega \int_0^\infty d \lambda \, \frac{ \lambda \, \sinh 2 \pi \lambda}{\cosh 2\pi \lambda -  \cos 2 \pi b} \log [1 - e^{-\omega(\lambda^2 + b^2 + m^2 a^2)}] \nn \\
 \beta_{1/2\,,0}(\omega) &:=&- \omega b \log [1 - e^{-\omega  m^2 a^2}] \\
 \beta_{1/2\,,D}(\omega) &:=& - 2 \omega \sum_{k=1}^{\lbrack b \rbrack} (b-k) \log[1 - e^{-\omega \left (-(k-b)^2 + b^2 + m^2 a^2 \right)}] \nn \,. 
\eeqa
Following the same steps as in the previous section, we can easily see that we can compare 
$\beta_{1/2}(\omega)$ with the flat space limit
\be
\beta^{\rm flat}_{1/2}(\omega ) = - \omega\, b \,\log(1-e^{-\omega m^2 a^2}) -
2\, \omega\, b \, \sum_{k=1}^{k_{max} \rightarrow \infty} \log [ 1 - e^{- \omega (2 b k + m^2 a^2)}] \,.
\label{H2SpinorBetalimit}
\ee
Taking the ratio of these quantities, we define
\be
\gamma_{1/2}(\omega) = \frac{\beta_{1/2}(\omega)}{\beta^{\rm flat}_{1/2}(\omega )} \,,
\ee
If the transverse magnetic field is absent, then, as in the case of the scalar field, there are no discrete energy states and we have
\be
\gamma_{1/2}(\omega, b = 0) = - \frac{12}{\pi^2} \omega 
\int_0^\infty d \lambda \, \lambda \, (\coth  \pi \lambda)\, \log [1 - e^{-\omega \lambda^2}]
\ee

\begin{figure}[!htb]\centering
	\begin{minipage}[t]{0.5\textwidth}
		\centering
		\includegraphics[width=1\textwidth]{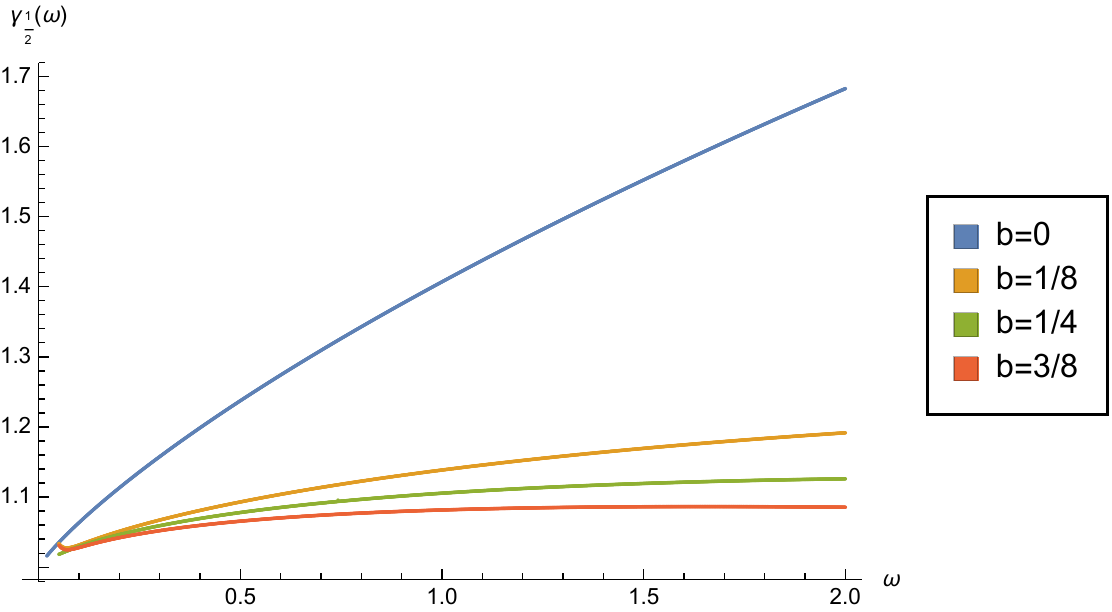}
		\caption{$\gamma_{1/2}$ versus $\omega$.} 
		\label{H^2gamma_half1}
	\end{minipage}%
	\begin{minipage}[t]{0.5\textwidth}
		\centering	
		\includegraphics[width=1\textwidth]{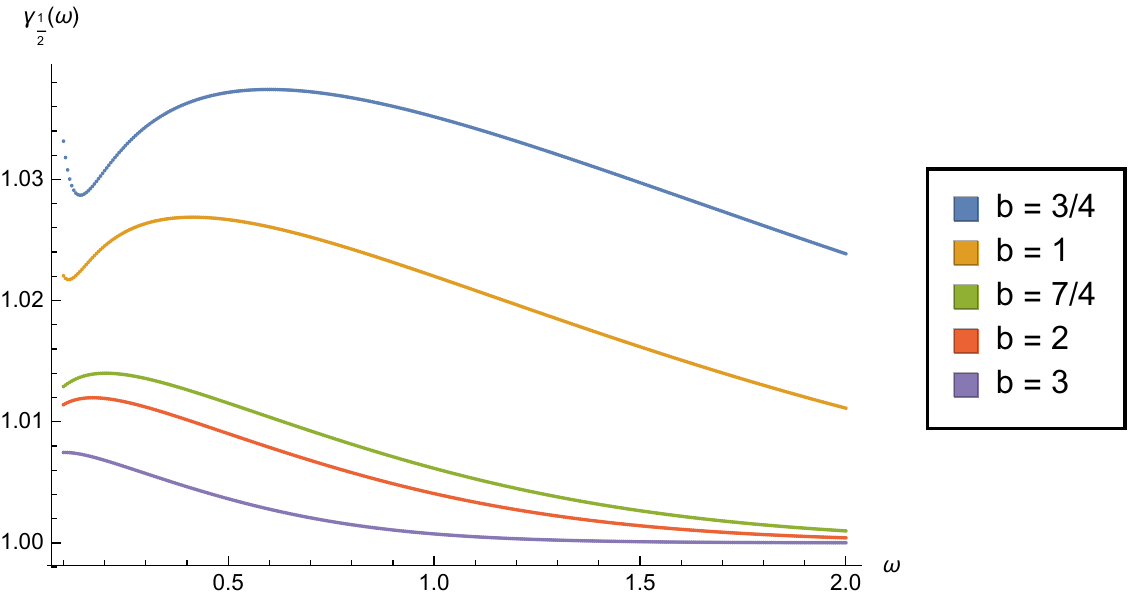}
		\caption{$\gamma_{1/2}$ versus $\omega$.} 
		\label{H^2gamma_half2}
	\end{minipage}
\end{figure}

The integrals over $\lambda$ can be done numerically to
graph out the profiles of $\gamma_{1/2} (\omega)$ for different values of the magnetic field $b$.
These are shown in Figs.\,\ref{H^2gamma_half1} and \ref{H^2gamma_half2}.
The graphs make it clear that at zero magnetic field, there is an increase in the pair production rate over and above what is obtained in the flat case. The underlying reason for this result is the fixed non-zero value of the density of states at small values of $\lambda$, since
$\lambda \coth \pi \lambda \rightarrow (d\lambda /\pi)$ as $\lambda \rightarrow 0$
compared to $dk k$ for flat space.
This allows for comparatively more particles to be accommodated at energies $\lambda^2/a^2 \approx 0$, i.e., almost without any energy cost. Once the $b$-field is switched on, there is always a zero-energy discrete state in the 
 spectrum with density $\frac{b}{2 \pi a^2}$, which essentially leads to the same enhancement effect as in the flat case. We note that the function $\gamma_{1/2}(\omega)$ contains the contribution of this zero mode term both in the numerator and the denominator, therefore it
 becomes rather insensitive to it at large $\omega$.
 Thus $\gamma_{1/2}(\omega)$ is basically controlled by
the curvature. With increasing magnetic field, we see that $\gamma_{1/2}(\omega)$ 
tends back to one at large $\omega$, 
meaning that the larger magnetic fields act to diminish the effect of curvature.

\begin{figure}[!htb]
	\centering
	\scalebox{.8}{\includegraphics[width=0.6\linewidth,height=0.25\textheight]{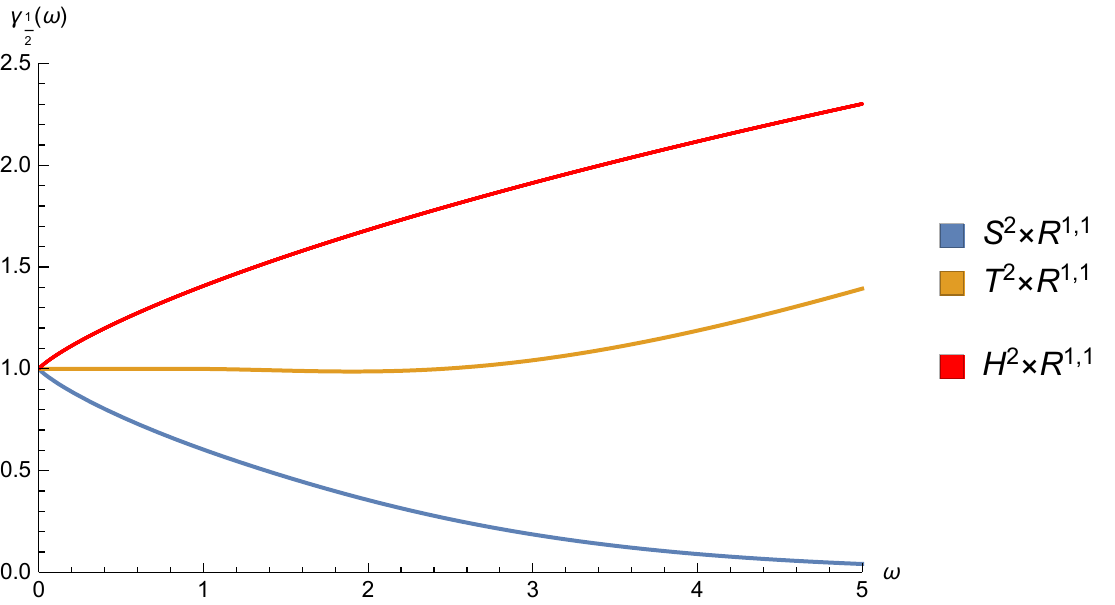}}
	\caption{Comparison of $\gamma_{1/2}$ at zero magnetic field for the torus, sphere and hyperboloid.}
	\label{gamma_0N=12}
\end{figure}

We close this section by giving a comparison of $\gamma_{1/2}(\omega)$ for the three cases of
torus, sphere and the hyperboloid for the case of zero magnetic
field in Fig.\,\ref{gamma_0N=12}.

\section{Discussion}

We have analyzed the pair production rates for
spin-zero and spin-$\half$ particles on spaces of the form
$M \times {\mathbb R}^{1,1}$ for $M = {\mathbb R}^2, \, T^2,\, S^2, \, H^2$.
We have also considered having a uniform background magnetic field
on $M$.
These cases allow us to compare the effects of curvature and magnetic
fields on the pair production rates. The analysis can be 
carried out using the representation theory for the appropriate isometry groups.
Our approach, while somewhat involved (particularly for $H^2$ where the principal series
play a significant role), 
does lead to explicit analytical formulae.
These results correspond to the evaluation of the
relevant determinants in integrating out the charged matter fields, or, equivalently
to one loop in matter fields.

There is a clear distinction which emerges for spin zero versus spin-$\half$.
On flat Minkowski space, a background magnetic field suppresses pair production
for the case of zero spin, since the produced pairs have to go into a Landau level and
there is a nonzero energy cost for this.
For spin-$\half$, the Dirac operator has a zero mode due to the 
magnetic moment coupling and hence there is an enhancement effect, with
an infrared divergence when the mass of the particle goes to zero.
This is very different from the situation with no magnetic field. Recall that
the Schwinger result (for rate per unit volume)
does not have a divergence even when the mass of the
fermion is zero.

Comparing $S^2$ and $H^2$, we find that there is a contrast,
with one leading to enhancement
and the other to suppression. Interestingly, the spins are also affected differently.
Thus we get enhancement for spin-zero and suppression for
spin-$\half$ for the case of positive curvature ($S^2$), while
there is suppression for spin-zero and enhancement for spin-$\half$ for negative
curvature ($H^2$). The interplay between the zero modes and the degeneracy
factors plays a crucial role in the difference between these cases.
Clearly the spin-curvature coupling and statistics
have a profound effect. 

As mentioned in the introduction, part of the motivation for our analysis was also to see
the impact of such considerations for vector particles and possible implications
for a nonabelian gauge theory.
This will be taken up in a follow-up paper \cite{KKN2}.

\section*{Appendix: Spectrum of Laplace and Dirac Operators on $H^2$}
\def\theequation{A\arabic{equation}}
\setcounter{equation}{0}

The spectrum of the Laplace and the Dirac operators on $H^2$ 
with a uniform background magnetic field, first appeared long time ago in an article by Comtet and Houston \cite{Comtet:1984mm} 
 and worked out in detail by Comtet in a subsequent paper \cite{Comtet:1986ki}. They are 
 encountered in the modern literature rather infrequently, so it is 
 useful to have a brief account of this to make the present paper self-contained. 

Since $H^2$ can be viewed as the coset space
\be
H^2 \equiv \frac{SU(1,1)}{U(1)} \,,
\ee
it is possible to employ the representation theory of $SU(1,1)$, 
or, equivalently, $SL(2,{\mathbb R})$, to obtain the spectrum of the Laplace and Dirac operators without reference to any particular coordinate system to describe $H^2$ and we will do so shortly. 
Nevertheless, it is useful to consider 
specific coordinate descriptions of the uniform magnetic field.
Following  \cite{Comtet:1984mm} , a convenient choice is to use the Poincaré coordinates, with which $H^2$ can be visualized as the upper half complex plane with the coordinates $z \equiv x + i y$, $y \geq 0$, and the metric
\be
ds^2 = \frac{a^2}{y^2} (d x^2 + dy^2) \,,
\ee
with the constant negative curvature $-\frac{2}{a^2}$.

The gauge potential
 and the corresponding field strength on $H^2$ can be given as the one-form $A = A_i dx^i = A_x dx + A_y dy$ and the two-form $F = d A$, respectively. Constant field strength on $H^2$ amounts to having $F$ proportional to the volume form on $H^2$, that is,
\be
F = \alpha \frac{a^2}{y^2} dx \wedge dy \,,
\label{FH2}
\ee
$\alpha$ being the constant of proportionality. In the Landau gauge, $(A_x, A_y) = (-\frac{b}{y}, 0)$, this takes the form
\be
F =  -\frac{b}{y^2} d x \wedge dy \,,
\ee
which gives the constant of proportionality in (\ref{FH2}) as $\alpha =  -\frac{b}{a^2}$, and $b$ can be used as a dimensionless parameter characterizing  the strength of the uniform magnetic field on $H^2$. Contrary to the case of the compact manifold $S^2$, there is no Dirac quantization condition on the magnetic field, therefore $b$ can be any real number.
Another useful coordinate system is to map $H^2$ to the unit disc in the complex
plane, with the metric and volume form
\beq
ds^2 = {d\bz\, dz \over (1- \bz z)^2}, \quad \quad
d\mu = {d\bz \wedge dz \over 2 i (1- \bz z)^2}
\label{h2metric}
\eeq
Again $F$ proportional to $d\mu$ would qualify as a uniform magnetic field.
There is also another set of coordinates where we map to
$\vert z\vert > 1$.

It is certainly  possible to express the Laplace and Dirac operators
with this uniform background magnetic field  in the coordinates give above.
Nevertheless group theoretical techniques are much more practical in obtaining
 the spectrum of these operators.
The generators $R_i \,,(i =1,2,3)$ of $SL(2,R)$, with the complex combinations $R_\pm = R_1 \pm i R_2$ satisfy the commutation relations
\be
\lbrack R_3 \,, R_\pm \rbrack =  \pm R_\pm \,, \quad \lbrack R_+ \,, R_- \rbrack = -2 R_3 \,.
\label{su11}
\ee
The quadratic Casimir operator for this group can be written as $R^2 := R_1^2 +R_2^2 -R_3^2$ and in complete analogy to the spherical case, the Laplace operator in the uniform magnetic field background can be expressed as 
\be
-D^2_{H^2} := \frac{1}{2 a^2} ( R_+ R_- + R_- R_+) = \frac{1}{a^2} (R^2 + R_3^2) \,,
\ee
We note the minus sign on the r.h.s of the second commutator in (\ref{su11}), compared to the $SU(2)$ commutation relations. This is reflected as the change of the sign before $R_3^2$ in the Casimir and subsequently in the Laplace operators. In order to compute the spectrum of this operator, we need to use the properties of unitary irreducible representations (UIRs) of $SL(2,R)$. The latter essentially splits in two parts, namely, the discrete series representations, which are semi-infinite dimensional, since they are bounded either from above or below, and the principal continuous series.\footnote {There is also the supplementary series UIR of $SL(2,{\mathbb R})$, but this does not arise in the computation of the spectrum of $-D^2_{H^2}$, therefore we do not discuss it here.} 

The discrete series UIRs of $SL(2,{\mathbb R})$ are characterized by a real number $\Lambda \geq \frac{1}{2}$, which is usually called the extremal weight of the UIR and in terms of this number, the eigenvalues of the Casimir operator are given as $R^2 \equiv - \Lambda (\Lambda -1)$. There are two equivalent representations corresponding to the same extremal weight, which are the discrete series bounded from above and below. Labeling the states in a UIR with the extremal weight $\Lambda$ and the eigenvalues $\Lambda + m$, $m=0,1,2,...$ of the generator $R_3$, we may explicitly express these representations as
\be
R^2 | \Lambda ,m \rangle = - \Lambda (\Lambda -1) | \Lambda ,m \rangle \,,\quad 
R_3 | \Lambda ,m \rangle =  \pm (\Lambda+ m) | \Lambda ,m \rangle \,,
\label{duirr}
\ee
where the representation with the upper sign for the $R_3$ eigenvalue has a lowest weight state and the one with the lower sign has a highest weight state and therefore bounded below and above, respectively.
The inner product for states in the representations bounded below 
is of the form
\beq
\braket{f|g}=  {2 \Lambda -1 \over \pi} \int {d\bz \wedge dz \over  2 i (1- \bz z)^2 } \, {\bar f} g,
\quad \vert z\vert < 1.
\eeq
The inner product for the states bounded above have a similar form,
\beq
\braket{f|g}=  {2 \Lambda -1 \over \pi} \int {d\bz \wedge dz \over  2 i ( \bz z -1 )^2 } \, {\bar f} g,
\quad \vert z\vert >1.
\eeq
Once we have chosen a parametrization of $H^2$ and a volume form,
only one of the two sets of representations will have
finite norm. Thus we can restrict to one of the two discrete sets of representations.
We will use those bounded below, so that, for us, $\Lambda >\half$.

The principal continuous series representations of $SL(2,R)$ are specified by the Casimir 
eigenvalue $\lambda^2 + \frac{1}{4}$, and the eigenvalue of $R_3$, which can be any real number and therefore this representation is not bounded either from above or 
below.\footnote { The extra $\frac{1}{4}$ means that
the eigenvalues never go down to zero. This is essentially the Breitenlohner-Freedman bound
\cite{BF}.} In particular, the harmonic functions on $H^2$ carry this representation. These features of $H^2$ are discussed at varying levels of detail and sophistication in the literature \cite{{Perelomov}, {Kitaev:2017hnr} }, but we will not dwell upon them as they are not necessary for our purposes in this article.

For the dynamics of charged particles on a plane subjected to a uniform perpendicular magnetic field as well as the corresponding spherical problem, which give the launching point for the discussion of quantum Hall effect on these geometries, semi-classical arguments immediately indicate that the particles move in circular orbits with cyclotron frequency proportional to the applied magnetic field (For a review, see for example \cite{Jain}) and there are infinite number of discrete energy levels in which the particles can be at, which are usually called the Landau levels. However, this picture no longer provides the complete description of the dynamics if the underlying space has negative curvature, which is the case for the present problem on $H^2$. For a given magnetic field on $H^2$, there are, in fact, only a finite number of discrete energy states, i.e. Landau levels, corresponding to the closed cyclotron orbits in the semi-classical description, the reason being essentially the constant negative curvature of $H^2$ acting against the formation of closed orbits. Therefore, the rest of the energy eigenstates are not quantized, but form a continuous spectrum \cite{{Comtet:1986ki},{MP}}. 

Without reference to the Poincaré coordinates or any other coordinate system for $H^2$, we may express the covariant derivatives on $H^2$ as $ D_\pm =   i {R_\pm / a}$. Commutator of the covariant derivatives is $\lbrack D_+ \,, D_- \rbrack = - 2 F = 2 \frac {b}{a^2}$ as usual and from the commutation relations of $R_\pm$, we infer that for the uniform magnetic field background we have to fix the eigenvalue of $R_3$ to be equal to $b$. Since there is no physical restriction over $b$ to be an integer, this means that $b$ labels not the UIRs of $U(1)$ in the coset description of $H^2$, but rather the UIRs of the universal cover $R$ of $U(1)$.

The generic representation of $SL(2,R)$ whose branching under the $R \simeq U(1)$ subgroup containing the UIR of the latter labeled by $b$ has the extremal weight
$\Lambda = b - k$ with $k \in {\mathbb Z}_+$. Therefore, the discrete part of the spectrum of the Laplacian is
\beqa
-D^2_{H^2} &=& \frac{1}{a^2}\left (-\Lambda (\Lambda -1) + R_3^2 \right) \nn \\
&=& \frac{1}{a^2}\left ( - (b-k)(b-k-1) + b^2 \right) \nn \\
&=& \frac{1}{a^2}\left ( -k(k+1) + 2 b k + b \right) \,, 
\label{dspech2}
\eeqa 
where $k=0,1,2,...$ labels the Landau levels (LLs). The ground state, $|b,0 \rangle$, is specified by taking $k=0$ and has the energy $\frac{b}{a^2}$. From the representation theory, the condition $\lambda = b-k \geq \frac{1}{2}$ has to be fulfilled and this gives $k \leq \lbrack b - \frac{1}{2} \rbrack$. This  means that, for a given value of $b$, there are only as many LLs as allowed by this inequality and they are labeled by the integers $k$. In particular, there are no LLs at all for $0 \leq b < \frac{1}{2}$.

Let us also remark that, we have used the UIR in (\ref{duirr}) with the upper sign, i.e. the one bounded from below, this fact can be concretely expressed as the lowering operator $R_-$ annihilating the lowest weight state: $R_-|b,0 \rangle = 0$.

Proceeding in the same manner, we see that the continuous part of the spectrum has the eigenvalues given by 
\be
-D^2_{H^2} = \frac{1}{a^2}\left ( \lambda^2 + \frac{1}{4} + b^2 \right ) \,,
\label{cspech2}
\ee
and it is readily observed from (\ref{dspech2}) and (\ref{cspech2}) that at any given value of $b$, the continuous part of the spectrum has larger eigenvalues than the discrete part as one would also expect from the preceding remarks on the semi-classical treatment of the problem. Detailed discussion of these features may be found in \cite{Comtet:1986ki}.

The density of the quantum states in the discrete and the principal continuous series representations are computed in the literature. Since, the derivations of these results are a bit long, we simply state these formulas and direct the reader to the original references in the literature, which are
 \cite{Comtet:1984mm} 
and \cite{Comtet:1986ki} , while for a recent extensive account based on the UIR theory of $SL(2,{\mathbb R})$, \cite{Kitaev:2017hnr}  can be consulted. For the discrete series representations, $SU(1,1)\simeq SL(2,R)$ 
we can use the coherent state basis \cite{Perelomov} to obtain the normalization of the energy eigenstates and this leads to the result
\be
\rho_{b}^{(0)}(k) = \frac{1}{2\pi a^2}\left(b - k-\frac{1}{2} \right) \,, \quad b > \frac{1}{2} \,,
\ee
Using the orthogonality property of the Wigner ${\cal D}$-functions for $SL(2,R)$, normalization of the energy eigenstates for the continuous part of the spectrum can be determined and this leads to the density of states given as 
\be
\rho_{b}^{(0)}(\lambda) = \frac{1}{2\pi a^2}\frac{ \lambda \, \sinh 2 \pi \lambda}{\cosh 2\pi \lambda +  \cos 2 \pi b} \,, \quad b \neq {\mathbb Z} + \frac{1}{2} \,.
\label{cdensityh2}
\ee
As $\lambda \rightarrow 0$, the density $\rho_{b}(\lambda) \rightarrow 0$, 
when half-integral values of $b$ are excluded and at half-integral values of $b$, the 
$\lambda \rightarrow 0$ limit of $\rho_{b}(\lambda)$ is $\frac{1}{2 \pi^2 a^2}$, although values of $b$ arbitrarily close to half-integers are allowed. (\ref{cdensityh2}) can be conceived as the Plancherel measure for the sections of the $U(1)$-bundle over $SL(2,{\mathbb R})$ and for $b=0$ it takes the form \cite{{Perelomov},{Kitaev:2017hnr}}
\be
\rho_{b=0}(\lambda) = \frac{1}{2\pi a^2} \lambda \tanh \pi \lambda \,.
\ee

The square of the Dirac operator on $H^2$ can be expressed as 
\beqa
-\slashed{D}^2 = - (\gamma\cdot D)^2 &=& {1\over a^2} \left[( R_1^2 + R_2^2) + \sigma_3 R_3 \right] \nn  \\
&= & {1\over a^2} \left[ R^2 + R_3^2 + \sigma_3 R_3 \right] \,.
\label{Sp-DirH2}
\eeqa
where the sign in front of the Zeeman-type term is flipped compared to the spherical case (\ref{Sp-Dir2}), as a reflection of the sign of the $R_+$, $R_-$ commutator in (\ref{su11}). The discrete part of the spectrum for the spin-up component (indicated by a subscript $+$ below)
follows from writing
\be
\Lambda = b - \frac{1}{2} -k \,, \quad \Lambda > \frac{1}{2} \,, \quad R_3 = b - \frac{1}{2} \,,  
\ee
which yields
\beqa
{\cal\mbox{Spec}}_D(-\slashed{D}^2_{+}) &=&  {1\over a^2}  \left [ -\Lambda (\Lambda -1) + R_3^2 + R_3 \right ] \nn \\
&=& {1\over a^2}  \left [- (b - \frac{1}{2} -k) (b - \frac{1}{2} -k - 1) + (b - \frac{1}{2})^2 + (b - \frac{1}{2}) \right ] \nn \\
&=& {1\over a^2} [- k^2 -2k + 2 b k + 2b - 1 ] \,, \quad k \leq [b-1] \,.
\eeqa
while for the spin down component, we have
\be
\Lambda = b + \frac{1}{2} -k \,, \quad \Lambda > \frac{1}{2}  \,, \quad R_3 = b + \frac{1}{2} \,, 
\ee
and this yields
\beqa
{\cal\mbox{Spec}}_D(-\slashed{D}^2_{-})
&=&  {1\over a^2}  \left [ -\Lambda (\Lambda -1) + R_3^2 - R_3 \right ] \nn \\
&=& {1\over a^2}  \left [- (b + \frac{1}{2} -k) (b + \frac{1}{2} -k - 1) + (b + \frac{1}{2})^2 -(b + \frac{1}{2}) \right ] \nn \\
&=& {1\over a^2}   [- k^2 + 2 b k ]  \,, \quad k \leq [b] \,,
\eeqa
For the continuous part of the spectrum, using the principal series UIR, we find the same spectrum for both the spin-up and the down components
\beqa
{\cal\mbox{Spec}}_C(-\slashed{D}^2_{\pm}) &=&  {1\over a^2}  \left [ \lambda^2 + \frac{1}{4} + ( b \pm \frac{1}{2})^2 \mp (b \pm \frac{1}{2}) \right ] \nn \\
&=& {1\over a^2}  \left [\lambda^2 + b^2 \right ] \,.
\eeqa
Similar considerations using the normalization for the coherent states and Wigner ${\cal D}$-functions for the spinor case leads to the densities
\beqar
\rho_{b}^{(1/2)}(k) &=& \frac{1}{2\pi a^2}\left(b - k \right) \,, \quad k \leq [b]
\nonumber\\
\rho_{b}^{(1/2)}(\lambda) &=& \frac{1}{2\pi a^2}\frac{ \lambda \, \sinh 2 \pi \lambda}{\cosh 2\pi \lambda -  \cos 2 \pi b} \,, \quad b \neq {\mathbb Z} + \frac{1}{2} \,.
\eeqar
In particular, for $b=0$ this takes the form
\be
\rho_{b=0}^{(1/2)}(\lambda) = \frac{1}{2\pi a^2} \lambda \coth \pi \lambda \,.
\ee

\vskip 2em

\noindent{\bf \large Acknowledgments}

\vskip 1em

\noindent S.K.'s work was carried out during his sabbatical stay at the physics department of CCNY of CUNY and he thanks V.P. Nair and D. Karabali for the warm hospitality at CCNY and the metropolitan area. S.K. also acknowledges the financial support of the Turkish Fulbright Commission under the visiting scholar program.  
The work of VPN was supported in part by the U.S.\ National Science
Foundation grant PHY-1820721.vDK and VPN acknowledge the support of PSC-CUNY awards.,

\vskip 1em

\end{document}